\documentclass{article}%
\usepackage{amsmath}
\usepackage{amsfonts}
\usepackage{amssymb}
\usepackage{graphicx}%
\providecommand{\U}[1]{\protect\rule{.1in}{.1in}}

\begin{document}

\title{The Entropic Dynamics of Spin\ }
\author{Ariel Caticha and Nicholas Carrara\\{\small Department of Physics, University at Albany-SUNY, }\\{\small Albany, NY 12222, USA.}}
\date{}
\maketitle

\begin{abstract}
In the Entropic Dynamics (ED) approach the essence of quantum theory lies in
its probabilistic nature while the Hilbert space structure plays a secondary
and ultimately optional role. The dynamics of probability distributions is
driven by the maximization of an entropy subject to constraints that carry the
relevant physical information --- directionality, correlations, gauge
interactions, etc. The challenge is to identify those constraints and to
establish a criterion for how the constraints themselves are updated. In this
paper the ED framework is extended to describe a spin-1/2 point particle. In
ED spin is neither modelled as a rotating body, nor through the motion of a
point particle; it is an epistemic property of the wave function. The
constraint that reflects the peculiar rotational properties of spin is most
effectively expressed in the language of geometric algebra. The updating of
all constraints is carried out in a way that stresses the central importance
of symmetry principles. First we identify the appropriate symplectic and
metric structures in the phase space of probabilities, their conjugate
momenta, and the spin variables. This construction yields a derivation of the
Fubini-Study metric for a spin-1/2 particle which highlights its deep
connection to information geometry. Then we construct an ED that preserves
both the symplectic structure (a Hamiltonian flow) and the metric structure (a
Killing flow). We show that generic Hamiltonian-Killing flows are linear in
the wave function. Imposing further that the Hamiltonian be the generator of
an entropic evolution in time leads to an entropic dynamics described by the
Pauli equation. We conclude with a discussion of the new interpretation of the
formalism which yields a physical picture that is significantly different from
that provided by other interpretations.

\end{abstract}

\section{Introduction}

Quantum Mechanics (QM) has been commonly conceived as a generalization of
classical mechanics with an added stochastic element. A Hilbert space
structure was adopted as the foundation of the theory while the probabilistic
structure was reluctantly included in order to handle those peculiar physical
processes called measurements \cite{Dirac 1930}\cite{von Neumann 1955}.
Unfortunately, the dynamical and the probabilistic aspects of quantum theory
turned out to be incompatible with each other causing all sorts of conceptual
difficulties and endless controversy.\footnote{Excellent reviews with extended
references to the literature are given in \emph{e.g.} \cite{Bell
1990}-\cite{Leifer 2014}.} Prominent among these difficulties is the question
of the interpretation of the quantum state itself: does the wave function
represent the actual state of the system (its \emph{ontic} state) or does it
represent our knowledge about the system (a merely \emph{epistemic}
state)?\footnote{A note on terminology: a concept is referred to as `ontic'
when it describes something that is supposed to be real, to exist out there
independently of any observer. A concept is referred to as `epistemic' when it
is related to the state of knowledge, opinion, or belief of an agent. (Here we
are only concerned with idealized rational agents.) Some examples of epistemic
quantities are probabilities and entropies. An important point is that the
distinction ontic/epistemic is not the same as the distinction
objective/subjective. For example, probabilities are fully epistemic --- they
are tools for reasoning with incomplete information --- but they can lie
anywhere in the spectrum from being completely subjective (two different
agents can have different beliefs) to being completely objective. In QM, for
example, probabilities are epistemic and objective. Indeed, at the
non-relativistic level anyone who computes probabilities that disagree with QM
will be led to experimental predictions that are demonstrably wrong. We will
say that the wave function $\Psi$, which is fully epistemic and objective,
represents a \textquotedblleft physical\textquotedblright\ state when it
represents information about an actual \textquotedblleft
physical\textquotedblright\ situation.}

Entropic Dynamics (ED) resolves these difficulties by placing the
probabilistic structure at the very foundation of the theory \cite{Caticha
2010}\cite{Caticha 2019}. Hilbert spaces do eventually appear as an emergent
or derived structure that plays the secondary role of a useful calculational
tool. In the ED approach QM\ is derived as an application of entropic methods
of inference.\footnote{The principle of maximum entropy as a method for
inference can be traced to the pioneering work of E. T. Jaynes \cite{Jaynes
1983}-\cite{Jaynes 2003}. For a pedagogical overview of Bayesian and entropic
inference and further references see \cite{Caticha 2020}.} The first concern
is to achieve ontological clarity: What is real? What are the probabilities
about? Once this issue is settled one can proceed to introduce probabilities
following the standard practice of maximizing an entropy subject to
constraints. The concern here is to identify the right constraints since it is
through them that the information that is relevant to the particular physics
problem is introduced.

In ED the central question is not how to obtain probabilities from a wave
function, how to derive the Born rule, but the reverse: given a purely
probabilistic foundation how do wave functions involving complex numbers
arise? Eventually a wave function $\Psi$ is introduced and it turns out to be
a fully epistemic concept. The demand for ontological clarity places
additional constraints on the theory. Once we decide to formulate a model in
which the ontic state of the system is defined by the positions of the
constituent particles then all other dynamical variables such as energy,
momentum, and angular momentum are necessarily epistemic. These are not
properties of the particles but properties of the wave function. Prior to a
measurement these variables need not have definite values and their values are
\textquotedblleft created\textquotedblright\ by the process of measurement.

The ED approach has been applied to a variety of topics in quantum theory: the
quantum measurement problem \cite{Johnson Caticha 2011}\cite{Vanslette Caticha
2016}; momentum and uncertainty relations \cite{Nawaz Caticha 2011}%
\cite{Bartolomeo Caticha 2016}; the Bohmian limit \cite{Bartolomeo Caticha
2016}\cite{Bartolomeo Caticha 2015} and the classical limit \cite{Demme
Caticha 2016}; extensions to curved spaces \cite{Nawaz et al 2015} and to
relativistic fields \cite{Ipek Caticha 2014}\cite{Ipek Abedi Caticha
2018}\cite{Ipek Caticha 2020}. In this paper the ED framework is extended to
describe the quantum mechanics of spin. Here we discuss a single spin-1/2
particle; in subsequent publications we will discuss multiparticle systems and
the Dirac equation.

Ever since Pauli's introduction of \textquotedblleft a peculiar, not
classically describable two-valuedness\textquotedblright\ \cite{Jammer 1966}
the nature of spin --- an intrinsic, or internal, or inherent angular momentum
--- has been the subject of numerous studies. The interpretations range from
the outright denial of the possibility of visualizable models to a wide
variety of alternatives. One possibility is that the particle is a small
spinning body described either by Bohmian mechanics \cite{Bohm Hiley
1993}\cite{Holland 1993}, or by stochastic mechanics \cite{Dankel 1970}, or
even by an entropic dynamics \cite{Nawaz 2014}. There are also hydrodynamic
models in which a particle-like inhomogeneity is carried along by a fluid
composed of rigid spinning bodies \cite{Bohm et al 1955}\cite{Takabayasi
1983}. Yet another approach is to attribute the spin to the motion of the
point particle guided by a pilot wave represented either by a Pauli wave
function \cite{Bell 1971}, or by a Dirac wave function \cite{Bohm Hiley
1993}\cite{Ohanian 1986}, or by a real spinor function expressed in the
language of geometric algebra \cite{Hestenes 1990}\cite{Doran et al 1996}. In
all of these models \emph{spin is an ontic variable} attributed either to the
spinning particle, or to the particle's motion, or to an ontic wave function.
All these models have been met with some degree of success, but they share
those characteristics of the hydrodynamic, Bohmian, or stochastic models, that
have traditionally stood in the way of their broader acceptance. In
particular, they provide no insights into the question of why does QM take the
particular form it does.

Our purpose here is to formulate a nonrelativistic ED model in which
\emph{spin is a purely epistemic property}. We deal with a point particle;
position is the only ontic variable and spin is a property of the wave
function. The end product is a new theory of spin. The entropic framework
leads, of course, to the same well-known equation derived by Pauli long ago
--- otherwise the theory presented here would be wrong --- but the
interpretation of \emph{what spin is} and \emph{how spin affects the motion of
a particle} is new.

As mentioned above ED consists in the entropic updating of probabilities with
information supplied by constraints. The challenge is to identify those
constraints that correspond to spin and, in particular, to establish a clear
criterion for how the constraints themselves are updated. The argument here
generalizes our previous discussion for scalar particles \cite{Caticha 2019}.
The idea, which stresses the central importance of symmetry principles when
implemented in conjunction with concepts of information geometry, is to
identify the appropriate symplectic and metric structures in the phase space
of probabilities, their conjugate momenta, and the spin
variables.\footnote{There is a vast literature on the geometric and symplectic
structure of quantum mechanics \cite{Hermann 1965}-\cite{Elze 2012} including
various aspects of the connection to information geometry \cite{Wootters
1981}-\cite{Molitor 2015} and \cite{Caticha 2019}.} The construction yields a
derivation of the Fubini-Study metric for a spin-1/2 particle which highlights
its deep connection to information geometry. A welcome by-product is that the
joint presence of a symplectic and a metric structure leads to a complex
structure which in the case of spin receives a particularly elegant geometric
algebra interpretation: the imaginary unit coincides with the pseudoscalar of
Euclidean space \cite{Hestenes 1966}\cite{Doran Lasenby 2003}. Once these
structures are in place we can formulate the desired geometric criterion for
updating constraints: we construct an ED that preserves both the symplectic
structure (a Hamiltonian flow) and the metric structure (a Killing flow). The
result of these Hamiltonian-Killing flows is an Entropic Dynamics described by
the linear Pauli equation.

The paper is structured as follows. In Section 2 we introduce the main ideas
behind ED. These include the choice of constraints that allow us to assign the
probability of a short step and the introduction of the notion of entropic
time as a tool to keep track of how these short steps accumulate. The
geometric criteria for updating the constraints so that the dynamics preserves
the appropriate symplectic and metric structures are discussed in Sections 3
and 4 respectively. In Section 5 we study the Hamilton-Killing flows and show
they lead to linear evolution and, in Section 6 the Pauli equation is derived.
Hilbert space is introduced in Section 7 as a tool that offers significant
calculational advantages but carries no deeper ontic or epistemic
significance. Then, in Section 8, we discuss the new interpretation of the
formalism which yields a physical picture that differs significantly from the
standard Copenhagen, Bohmian, or even the Hestenes interpretations. Section 9
summarizes results and includes some final comments. To streamline the
presentation some steps in the proofs have been relegated to the appendices.
Appendix A contains a brief review of the Geometric Algebra approach to
spinors that serves to establish the notation and collect in one place various
relevant formulas.

\section{Entropic dynamics of a spin-1/2 particle}

Our goal is to study the evolution of the probability $\rho(x)$ of the
position $x=(x^{1},x^{2},x^{3})\in\mathbf{X}$ of a point particle that lives
in a three-dimensional Euclidean space $\mathbf{X}$. In contrast to the
standard Copenhagen interpretation in our model the particle has a definite
position at all times. Furthermore, in contrast to other approaches to spin,
we deal with a point particle and not with a small rotating body. In the ED
approach spin is a property of the motion and not a property of the particle's
internal structure.

The first step in defining the dynamics is to find the probability
$P(x^{\prime}|x)$ that the particle moves from a point $x$ to a neighboring
point $x^{\prime}$. This is done by maximizing the entropy,\footnote{To
simplify the notation in multidimensional integrals we shall write $dx$ rather
than $d^{3}x$.}%
\begin{equation}
\mathcal{S}[P,Q]=-\int dx^{\prime}\,P(x^{\prime}|x)\log\frac{P(x^{\prime}%
|x)}{Q(x^{\prime}|x)}~,~ \label{entropy a}%
\end{equation}
of $P(x^{\prime}|x)$ relative to a prior $Q(x^{\prime}|x)$ and subject to the
appropriate constraints that incorporate the relevant information about the
motion. It is through the prior and the constraints that the \textquotedblleft
physical\textquotedblright\ information is introduced.

\subsection{The prior}

The main dynamical assumption is that \emph{the particle follows a continuous
path}. This allows us to analyze a generic motion as the accumulation of many
infinitesimally short steps. We shall adopt a prior that forces the particle
to take short steps but is otherwise maximally uninformative. Such a prior can
itself be derived using the principle of maximum entropy. Indeed, maximize
\begin{equation}
S[Q,\mu]=-\int dx^{\prime}\,Q(x^{\prime}|x)\log\frac{Q(x^{\prime}|x)}%
{\mu(x^{\prime}|x)}~,
\end{equation}
relative to the uniform measure $\mu$,\footnote{In Cartesian coordinates the
uniform measure $\mu$ is a constant that may be ignored.} subject to
normalization, and subject to a constraint that imposes short steps and
implements the translational and rotational invariance of space. The
constraint is
\begin{equation}
\langle\delta_{ab}\Delta x^{a}\Delta x^{b}\rangle=\kappa~,\quad
\end{equation}
where $\Delta x^{a}=x^{\prime a}-x^{a}$ and $\kappa$ is a small constant. The
result is a Gaussian,
\begin{equation}
Q(x^{\prime}|x)\propto\exp\left(  -\frac{1}{2}\alpha\delta_{ab}\Delta
x^{a}\Delta x^{b}\right)  ~, \label{prior}%
\end{equation}
where $\alpha$ is a Lagrange multiplier that will eventually be taken to
infinity in order to enforce the fact that the steps are meant to be
infinitesimally short.\footnote{Since the maximization of an entropy in the
context of a dynamical non-equilibrium situation might not be familiar to some
readers it might be useful to point out that the process yields the broadest
distribution consistent with the imposed constraint. In our present problem
due to the constraint being very narrow the maximum entropy distribution is
forced to be extremely sharp.}

\subsection{The constraints}

The distribution (\ref{prior}) leads to a diffusion process that is isotropic.
We seek to generate a motion that exhibits directionality, the interaction
with electromagnetic fields, and the peculiar rotational properties attributed
to spin. This information is incorporated by imposing additional constraints
on the expected displacement,
\begin{equation}
\langle\Delta x^{a}\rangle=\int dx^{\prime}\,P(x^{\prime}|x)\Delta x^{a}.
\end{equation}
One can reasonably expect that the dynamics of particles with spin should, at
least in those situations when the spin happens to be uniform and constant,
resemble the dynamics of scalar particles. Accordingly, we shall construct the
ED of a particle with spin in close analogy to the ED of scalar particles
\cite{Caticha 2019}.

\paragraph*{The drift potential constraint ---}

The physical information that the motion can be directional is introduced
through a constraint expressed in terms of a drift potential $\varphi(x)$.
{}As was already the case for scalar particles, that a variable such as
$\varphi$ must be introduced becomes clear once one realizes that any
Hamiltonian dynamics of probabilities $\rho$ must inevitably involve the
canonically conjugate momenta. The constraint consists of requiring that the
expected displacement $\langle\Delta x^{a}\rangle$ along the gradient of
$\varphi$ be a small quantity $\kappa^{\prime}$,\
\begin{equation}
\langle\Delta x^{a}\rangle\partial_{a}\varphi=\kappa^{\prime}~.
\label{constraint phi}%
\end{equation}
The value of the quantity $\kappa^{\prime}$ could be given directly, but as we
shall see below, it turns out to be much more convenient to do so indirectly.
We shall follow the practice that is standard in statistical mechanics where
rather than specifying the expected values one specifies the corresponding
Lagrange multipliers.

\paragraph{The gauge constraint ---}

Just as for scalar particles, to incorporate the effect of an external
electromagnetic field we impose that the expected displacement $\langle\Delta
x^{a}\rangle$ satisfy%
\begin{equation}
\langle\Delta x^{a}\rangle A_{a}(x)=\kappa^{\prime\prime}~,~
\label{constraint A}%
\end{equation}
where $A_{a}(x)$ is the electromagnetic vector potential and $\kappa
^{\prime\prime}$ is yet another small quantity to be (indirectly) specified below.

\paragraph*{The spin constraint ---}

The framework of Geometric Algebra is particularly well suited for the study
of spin because it provides an efficient way to represent rotations (see
Appendix A).\footnote{GA is a Clifford algebra with an added geometric
interpretation: it is a Clifford algebra in which the relevant bilinear form
is a metric on the underlying vector space.} Let the three orthonormal vectors
$\vec{e}_{1}$, $\vec{e}_{2}$, and $\vec{e}_{3}$ be the standard basis for
Euclidean space --- the laboratory basis. The result of rotating the standard
frame by different amounts at different points $\vec{x}$ in space is a
\emph{spin frame or spin triad} \emph{field}, given by
\begin{equation}
\vec{s}_{k}(\vec{x})=U(\vec{x})\vec{e}_{k}U^{\dag}(\vec{x})~,
\label{spin frame}%
\end{equation}
where the \emph{rotor} $U(\vec{x})$ is a unitary quaternion --- a element of
the algebra that has even grade and unit magnitude (Appendix A.2.1). In terms
of the local Euler angles $\{\theta(\vec{x}),\phi(\vec{x}),\chi(\vec{x})\}$
the rotor $U(\theta,\phi,\chi)$ (Appendix A.2.2) is
\begin{equation}
U(\theta,\phi,\chi)=U_{z}(\phi)U_{y}(\theta)U_{z}(\chi)=e^{-i\vec{e}_{3}%
\phi/2}e^{-i\vec{e}_{2}\theta/2}e^{-i\vec{e}_{3}\chi/2}~. \label{rotor a}%
\end{equation}
The angles $\theta$ and $\phi$ are the polar angles of $\vec{s}_{3}$ and the
angle $\chi$ describes the rotation of $\vec{s}_{1}$ and $\vec{s}_{2}$ about
the $\vec{s}_{3}$ axis. From now on we shall drop the label $3$ and the vector
$\vec{s}_{3}=\vec{s}$ will be called the spin vector,
\begin{equation}
\vec{s}=U\vec{e}_{3}U^{\dag}=\sin\theta\cos\phi\,\vec{e}_{1}+\sin\theta
\sin\phi\,\vec{e}_{2}+\cos\theta\,\vec{e}_{3}~. \label{spin vector a}%
\end{equation}

An entropic dynamics that reflects the rotational properties of spin 1/2 is
obtained by imposing a constraint involving \textquotedblleft
potentials\textquotedblright\ associated to the rotor field $U(\theta
,\phi,\chi)$ or equivalently to the spin frame field $\{\vec{s}_{1},\vec
{s}_{2},\vec{s}_{3}\}$. This spin constraint resembles (\ref{constraint phi})
but, instead of the gradient $\partial_{a}\varphi$, it involves the natural
spatial derivatives $\vec{\omega}_{a}$ associated to the spin frame. As one
moves from $\vec{x}$ to $\vec{x}+\delta\vec{x}$ the spin frame undergoes an
infinitesimal rotation $\delta\vec{\zeta}(\vec{x})$ by an angle $|\delta
\vec{\zeta}|$ about the axis $\vec{n}=\delta\vec{\zeta}/|\delta\vec{\zeta}|$.
This is given by eq.(\ref{omega b}) and (\ref{omega c}),\footnote{We could
write $\vec{\omega}_{a}=\partial_{a}\vec{\zeta}$. Quantities such as
$\partial_{a}\vec{\zeta}$, which are typical in discussions of non-holonomic
constraints, are not integrable: although $\partial_{a}\vec{\zeta}$ is a well
defined function of $\vec{x}$ it cannot be integrated to yield a vector field
$\vec{\zeta}(\vec{x})$.}
\begin{equation}
\delta\vec{\zeta}=\vec{\omega}_{a}\delta x^{a}\quad\text{where}\quad
~\vec{\omega}_{a}=2i(\partial_{a}U)U~.~
\end{equation}
Expressed in terms of Euler angles we have%
\begin{equation}
\vec{\omega}_{a}=\vec{e}_{3}\partial_{a}\phi+\vec{e}_{\phi}\partial_{a}%
\theta+\vec{s}\,\partial_{a}\chi\quad\text{with}\quad\vec{e}_{\phi}=-\sin
\phi\,\vec{e}_{1}+\cos\phi\,\vec{e}_{2}~,
\end{equation}
and the components of $\vec{\omega}_{a}$ along the spin vector $\vec{s}$ are
\begin{equation}
\vec{\omega}_{a}\cdot\vec{s}=\partial_{a}\chi+\cos\theta\,\partial_{a}\phi~.
\end{equation}
We are now ready to write the constraint due to spin:%

\begin{equation}
\langle\Delta x^{a}\rangle\vec{\omega}_{a}\cdot\vec{s}=\langle\Delta
x^{a}\rangle\left(  \partial_{a}\chi+\cos\theta\,\partial_{a}\phi\right)
=\kappa^{\prime\prime\prime}~, \label{constraint spin}%
\end{equation}
where $\kappa^{\prime\prime\prime}$ is yet another small quantity to be
specified below.

\subsection{The probability of a short step}

Maximizing the entropy (\ref{entropy a}) subject to the constraints
(\ref{constraint phi}, \ref{constraint A}, \ref{constraint spin}) plus
normalization leads to the transition probability%
\begin{equation}
P(x^{\prime}|x)\propto\exp\left[  -\frac{\alpha}{2}\,\Delta x^{a}\Delta
x^{b}\delta_{ab}+\alpha^{\prime}\Delta x^{a}\left[  \partial_{a}\varphi-\beta
A_{a}-\gamma\left(  \partial_{a}\chi+\cos\theta\,\partial_{a}\phi\right)
\right]  \right]  ~, \label{trans prob a}%
\end{equation}
where $\alpha^{\prime}$, $\beta$, and $\gamma$ are Lagrange multipliers. Since
$P(x^{\prime}|x)$ is Gaussian, it can be more conveniently rewritten as
\begin{equation}
P(x^{\prime}|x)\propto\exp[-\frac{\alpha}{2}\,\delta_{ab}(\Delta x^{a}%
-\Delta\bar{x}^{a})(\Delta x^{b}-\Delta\bar{x}^{b})]~. \label{trans prob b}%
\end{equation}
A generic displacement $\Delta x^{a}=x^{\prime a}-x^{a}$, can be written as
\begin{equation}
\Delta x^{a}=\Delta\bar{x}^{a}+\Delta w^{a}~, \label{delta x a}%
\end{equation}
where $\Delta\bar{x}^{a}=\left\langle \Delta x^{a}\right\rangle $ is the
expected drift,%
\begin{equation}
\left\langle \Delta x^{a}\right\rangle =\frac{\alpha^{\prime}}{\alpha}%
\delta^{ab}\left[  \partial_{b}\varphi-\beta A_{b}-\gamma(\partial_{b}%
\chi+\cos\theta\,\partial_{b}\phi)\right]  ~, \label{drift a}%
\end{equation}
and $\Delta w^{a}$ is a fluctuation,%
\begin{equation}
\left\langle \Delta w^{a}\right\rangle =0\quad\text{and}\quad\left\langle
\Delta w^{a}\Delta w^{b}\right\rangle =\frac{1}{\alpha}\delta^{ab}~.
\label{fluct a}%
\end{equation}

Just as for scalar particles the fact that the constraints
(\ref{constraint phi}) and (\ref{constraint A}) are not independent --- both
involve the same displacements $\langle\Delta x_{n}^{a}\rangle$ --- leads to a
gauge symmetry. As we can see in (\ref{trans prob a}) and (\ref{drift a})
where $\varphi$ and $A_{a}$ appear in the combination $\partial_{a}%
\varphi-\beta A_{a}$ which is invariant under the gauge transformations,
\begin{align}
A_{a}(x)  &  \rightarrow A_{a}^{\prime}(x)=A_{a}(x)+\partial_{a}\xi(x)~,\\
\varphi(x)  &  \rightarrow\varphi^{\prime}(x)=\varphi(x)+\beta\xi(x)~.
\end{align}

As we mentioned earlier the values of the as yet unspecified quantities
$\kappa^{\prime}$, $\kappa^{\prime\prime}$, and $\kappa^{\prime\prime\prime}$
in the constraints (\ref{constraint phi}), (\ref{constraint A}), and
(\ref{constraint spin}) could be given directly --- just state what
$\kappa^{\prime}$, $\kappa^{\prime\prime}$, and $\kappa^{\prime\prime\prime}$
are --- but it is much more convenient to do so indirectly by specifying the
corresponding Lagrange multipliers. In the next section we address the issue
of assigning values to the multipliers $\alpha$ and $\alpha^{\prime}$. Later,
in section \ref{metric in T*C+1} we shall argue that the value of $\gamma$
that ensures compatibility with the rotational properties of spinors is
$\gamma=1/2$, and in section \ref{H and P eq} we shall see that the multiplier
$\beta$ will turn out to be related to the electric charge $q$ through
$\beta=q/\hbar c$.

\subsection{Entropic time: instants and duration}

ED consists of a succession of short steps; in order to keep track of how
these small changes accumulate one introduces the notion of time. As proposed
in \cite{Caticha 2010} and \cite{Caticha 2019} the construction of entropic
time involves the concepts of `instant' and of `duration'. An instant is
specified by the probability distribution $\rho(x)$ and by the transition
probability $P(x^{\prime}|x)$ given in (\ref{trans prob a}). The distribution
$\rho(x)$ implicitly provides a criterion of simultaneity and the pair
$(\rho,P)$ conveys the information required for the construction of the next
instant. More explicitly, if the distribution $\rho_{t}(x)$\ refers to one
instant $t$, then the distribution $\rho_{t^{\prime}}(x^{\prime})$\ generated
by $P(x^{\prime}|x)$ through
\begin{equation}
\rho_{t^{\prime}}(x^{\prime})=%
{\textstyle\int}
dx\,P(x^{\prime}|x)\rho_{t}(x) \label{ED a}%
\end{equation}
defines what we mean by the \textquotedblleft next\textquotedblright\ instant
which we have arbitrarily labelled $t^{\prime}$. Iterating this process
defines the entropic dynamics.\footnote{As noted in \cite{Caticha
2010}\cite{Caticha 2019} the fact that $P(x^{\prime}|x)$ is derived using the
maximum entropy method introduces an arrow of time --- entropic time only
flows towards the future. Nevertheless, the resulting Schr\"{o}dinger or Pauli
equations exhibits the symmetry commonly referred to as time reversal or (more
appropriately) motion reversal.}

To complete the construction of the dynamics and of entropic time we must
specify the Lagrange multipliers $\alpha$ and $\alpha^{\prime}$ in
$P(x^{\prime}|x)$, and also the duration or interval $\Delta t=t^{\prime}-t$
between successive instants. This is accomplished by specifying the relation
between $\alpha$, $\alpha^{\prime}$, and $\Delta t$. The basic criterion is
convenience: \textquotedblleft duration is defined so that motion looks
simple\textquotedblright\ \cite{Caticha 2019}. For the non-relativistic regime
considered in this paper we adopt a choice that reflects the translational
symmetry of a Newtonian space and time: we choose $\alpha^{\prime}$ and
$\alpha$ to be independent of $x$ and $t$ so that time flows \textquotedblleft
equably everywhere and everywhen,\textquotedblright\ and we choose the ratio
$\alpha^{\prime}/\alpha\propto\Delta t$ so that eq.(\ref{drift a}) leads to a
well defined drift velocity. In Newtonian mechanics the prototype of a clock
is a free particle with a well-defined velocity --- equal times are
\textquotedblleft ticked\textquotedblright\ by equal distances. In ED the
prototype of a clock is a free particle too in the sense that for sufficiently
short times all particles are effectively free and the ratio $\alpha^{\prime
}/\alpha$ is chosen so that the particle has a well defined \emph{expected}
velocity.\footnote{See however eq.(\ref{velocity a}) below.} The
proportionality constant is written as $\hbar/m$,
\begin{equation}
\frac{\alpha^{\prime}}{\alpha}=\frac{\hbar}{m}\Delta t~, \label{alphas}%
\end{equation}
where $m$ will eventually be identified as the particle's mass and $\hbar$ is
a constant that fixes the units of $m$ relative to the units of time.

Having specified the ratio $\alpha^{\prime}/\alpha$ it remains to specify
$\alpha^{\prime}$. It turns out that different choices of $\alpha^{\prime}$
lead to different forms of dynamics at the sub-quantum level. The choice
$\alpha^{\prime}=\operatorname*{const}$ leads to jagged, non-differentiable
trajectories characteristic of Brownian paths \cite{Bartolomeo Caticha 2015}.
Alternatively, the choice $\alpha^{\prime}\varpropto1/\Delta t^{2}$ leads to
the much smoother trajectories that we have elsewhere called \textquotedblleft
Bohmian\textquotedblright\ \cite{Caticha 2019}. Whether Brownian or Bohmian,
the important point is that in the end \emph{the choice of }$\alpha^{\prime}%
$\emph{ does not matter: }all these different forms of sub-quantum dynamics
lead to the same theory at the quantum level. Here we shall explore the
particular sub-quantum ED described by $\alpha^{\prime}\propto1/\Delta t^{2}$.
We write the proportionality constant as $\hbar/\eta$,
\begin{equation}
\alpha^{\prime}=\frac{\hbar}{\eta\Delta t^{2}}\quad\text{so that}\quad
\alpha=\frac{m}{\eta\Delta t^{3}}~. \label{alpha prime}%
\end{equation}
To motivate this choice of $\alpha^{\prime}$ one only needs to note that the
corresponding transition probability (\ref{trans prob b}) takes the
particularly simple form
\begin{equation}
P(x^{\prime}|x)=\frac{1}{Z}\exp\left[  -\frac{m}{2\eta\Delta t}\delta
_{ab}\left(  \frac{\Delta x^{a}}{\Delta t}-v^{a}\right)  \left(  \frac{\Delta
x^{b}}{\Delta t}-v^{b}\right)  \right]  ~, \label{trans prob c}%
\end{equation}
where we used (\ref{drift a}) to define the drift velocity,
\begin{equation}
v_{a}=\frac{1}{m}\partial_{a}\Phi-\frac{q}{mc}A_{a}-\frac{\hbar\gamma}{m}%
\cos\theta\,\partial_{a}\phi~. \label{drift c}%
\end{equation}
In this expression the drift potential $\varphi$ and the Euler angle $\chi$
have been merged into the single variable
\begin{equation}
\Phi=\hbar(\varphi-\gamma\chi) \label{phase a}%
\end{equation}
which we will call \emph{the phase}\footnote{In the multiparticle case it will
not be possible to combine the drift potential $\varphi$ with the Euler angles
$\chi$ associated to different particles into a single phase variable $\Phi$.
They are independent degrees of freedom \cite{Carrara Caticha 2021}.} and we
set $\beta=q/\hbar c$. A generic displacement is then written as an expected
drift plus a fluctuation,
\begin{equation}
\Delta x^{a}=v^{a}\Delta t+\Delta w^{a}~, \label{delta x}%
\end{equation}
with the fluctuations $\Delta w^{a}$ given by
\begin{equation}
\langle\Delta w^{a}\rangle=0\quad\text{and}\quad\langle\Delta w^{a}\Delta
w^{b}\rangle=\frac{\eta}{m}\delta^{ab}\Delta t^{3}~,
\end{equation}
or
\begin{equation}
\left\langle \left(  \frac{\Delta x^{a}}{\Delta t}-v^{a}\right)  \left(
\frac{\Delta x^{b}}{\Delta t}-v^{b}\right)  \right\rangle =\frac{\eta}%
{m}\delta^{ab}\Delta t. \label{fluct c}%
\end{equation}
Since $\langle\Delta x^{a}\rangle\sim O(\Delta t)$ and $\Delta w^{a}\sim
O(\Delta t^{3/2})$ we see that for small $\Delta t$ the fluctuations are
negligible and the particle follows a smooth deterministic trajectory. Indeed,
unlike Brownian trajectories, the limit
\begin{equation}
\lim_{\Delta t\rightarrow0}\frac{\Delta x^{a}}{\Delta t}=v^{a}
\label{velocity a}%
\end{equation}
is well defined. Not only does the particle have a \emph{definite position},
it also has a \emph{definite velocity} given by the expected drift velocity
(\ref{drift c}). This is a feature that this version of ED shares in common
with Bohmian mechanics.

\subsection{The evolution equation}

The basic evolution equation (\ref{ED a}) with (\ref{trans prob c}) can be
written in differential form,
\begin{equation}
\partial_{t}\rho=-\partial_{a}(\rho v^{a})~. \label{ED b}%
\end{equation}
The proof follows closely the analogous derivation given in \cite{Caticha
2019} for the scalar case and need not be repeated here. Equation (\ref{ED b})
states that the probability evolves according to a continuity equation with
probability flux $\rho v^{a}$, and the current velocity coincides with the
drift velocity $v^{a}$, eq.(\ref{drift c}).\footnote{When $\alpha^{\prime
}=\operatorname*{const}$ and the sub-quantum dynamics is a Brownian motion the
evolution equation (\ref{ED c}) is a Fokker-Planck equation with a current
velocity\ that differs from the drift velocity (\ref{drift c}) by an osmotic
term $\varpropto\log\rho^{1/2}$.}

\paragraph*{Introducing the ensemble Hamiltonian ---}

For future use we note that the evolution equation (\ref{ED b}) can be
conveniently rewritten in the alternative form
\begin{equation}
\partial_{t}\rho=-\partial_{a}(\rho v^{a})=\frac{\delta\tilde{H}}{\delta\Phi
}~, \label{ED c}%
\end{equation}
for some suitably chosen functional $\tilde{H}=\tilde{H}[\rho,\Phi,\phi
,\theta]$. Using (\ref{drift c}) we see that eq.(\ref{ED c}) is a first order
differential equation that can be easily integrated to give
\begin{equation}
\tilde{H}=\int d^{3}x\,\rho\frac{1}{2}m\delta_{ab}v^{a}v^{b}+F[\rho
,\phi,\theta]~, \label{Hamiltonian a}%
\end{equation}
where the integration constant $F[\rho,\phi,\theta]$ is some unspecified
functional of $\{\rho,\theta,\phi\}$. And, indeed, we can check that the
variation of $\tilde{H}$ with respect to $\delta\Phi$ with $\{\rho,\theta
,\phi\}$ fixed is
\begin{equation}
\delta\tilde{H}=\int dx\,\rho v^{a}\partial_{a}\,\delta\Phi=-\int
dx\partial_{a}(\rho v^{a})\,\delta\Phi~.
\end{equation}

With these results the ED of spin reaches a certain level of completion. Even
though the multipliers $\gamma$ and $\beta$ remain unspecified we have figured
out the short steps we ought to expect --- they are given by the transition
probability $P(x^{\prime}|x)$ in eq.(\ref{trans prob c}) --- and the concept
of time was introduced to keep track of how these small changes accumulate;
the net result is an evolution equation, eq.(\ref{ED c}), driven by various
\textquotedblleft potentials\textquotedblright\ --- the phase $\Phi$, the
angles $\theta$ and $\phi$\ associated to the spin, and the vector potential
$A_{a}$. The result is a consistent dynamics driven by constraints with some
peculiar rotational properties.

However, this ED is not yet quantum mechanics. So far the spin variables have
been treated as externally prescribed background fields but in a proper
quantum dynamics not only the probability $\rho$ but the phase $\Phi$ and the
spin variables themselves must also participate in the dynamics.

\section{Symplectic structure and Hamiltonian flows}

The ED that turns out to be physically relevant is achieved by requiring that
the constraints used to update the probability $\rho_{t}(x)$ are not kept
fixed. Instead they are themselves continuously updated in response to the
evolving $\rho_{t}$. Such an ED describes the coupled evolution of the four
variables $(\rho,\Phi,\theta,\phi)$ and is closely analogous to the ED\ of
scalar particles developed in \cite{Caticha 2019}: the criterion for updating
the constraints is that some suitably defined symplectic and Riemannian
structures are preserved by the dynamics.

\subsection{The epistemic phase space}

The symplectic structure we seek is chosen so that its preservation implies
the ED continuity equation (\ref{ED c}). This leads us to identify the phase
function $\Phi(x)$ introduced in (\ref{phase a}) as the momentum that is
canonically conjugate to the probability $\rho(x)$. In this section and the
next we tackle the question of how to handle the other two spin variables,
$\theta$ and $\phi$, and assign a value to the multiplier $\gamma$.

In the ED of a scalar particle we deal with two configuration spaces. One is
the \emph{ontic configuration space} $\mathbf{X}=\mathbf{R}^{3}$ of positions
and the other is the \emph{epistemic configuration space }%
or\emph{\ e-configuration space} of the normalized probabilities,
\begin{equation}
\mathbf{P}=\left\{  \rho\left\vert \rho(x)\geq0;\int dx\rho(x)=1\right.
\right\}  ~. \label{P}%
\end{equation}
For the single particle with spin the ontic configuration space $\mathbf{X}%
=\mathbf{R}^{3}$ remains unchanged but the epistemic configuration space will
be enlarged to include spin.

The choice of the configurational spin variable is suggested by an early work
of Kramers \cite{Kramers 1957} that was later adapted by Bohm, Schiller, and
Tiomno \cite{Bohm et al 1955}, and by Takabayasi \cite{Takabayasi 1983} to the
context of spin and the Pauli equation. Kramers observed that since the
orientation of a magnetic dipole with a fixed magnitude is defined by just two
degrees of freedom, the angles $\phi$ and $\theta$ that define its
orientation, then the minimal Hamiltonian dynamics --- minimal in the sense
that no additional variables are introduced --- is constructed by identifying
one of these variables as the coordinate and the other as its conjugate
momentum. We shall adopt the azimuthal angle $\phi$ of the spin vector
$\vec{s}$ for the coordinate variable so that the e-configuration space is
$\mathbf{C=P\times}\{\phi\}$. For the conjugate momentum we choose the
densitized $z$-component of the spin vector,
\begin{equation}
\rho_{s}=\frac{\hbar}{2}\rho\cos\theta~,
\end{equation}
which we shall call the spin density.\footnote{Alternative choices such as
$\theta$ or $\cos\theta$ represent canonical transformations which lead to the
same final dynamics. The factor $\hbar/2$ is included for later convenience.}

To formulate the dynamics we need a framework to study paths in the larger
space $\{\rho,\Phi,\phi,\rho_{s}\}$ which we will call the \emph{epistemic
phase space} or \emph{e-phase space}. Given a manifold such as $\mathbf{C}$
its cotangent bundle, $T^{\ast}\mathbf{C}$, is a geometric object that comes
automatically endowed with a rich geometric structure \cite{Kibble
1979}-\cite{Elze 2012}. Cotangent bundles are symplectic manifolds and this
provides a \textquotedblleft natural\textquotedblright\ criterion for
dynamical laws, namely those that happen to preserve a certain privileged
symplectic form. Thus, the natural criterion for updating constraints follows
from identifying the cotangent bundle\emph{\ }$T^{\ast}\mathbf{C}$ with
e-phase space $\{\rho,\Phi,\phi,\rho_{s}\}$,%
\begin{equation}
T^{\ast}\mathbf{C}=\{\rho,\Phi,\phi,\rho_{s}\}\mathbf{~.} \label{ephase space}%
\end{equation}

\paragraph*{Notation ---}

Our notation follows closely that adopted in \cite{Caticha 2019}. A point
$X=(\rho,\Phi,\phi,\rho_{s})$ will be labelled by its coordinates
\begin{equation}
X^{\alpha x}=(X^{1x},X^{2x},X^{3x},X^{4x})=\left(  \rho^{x},\Phi_{x},\phi
^{x},\rho_{sx}\right)  ~,
\end{equation}
where $(\rho^{x},\phi^{x})$ represent coordinates on the base manifold
$\mathbf{C}$ and $(\Phi_{x},\rho_{sx})$ represent coordinates on the space
$T^{\ast}\mathbf{C}_{\rho\phi}$ that is cotangent to $\mathbf{C}$ at the point
$(\rho,\phi)$. We shall use a composite index $\alpha x$ with the Greek index
$\alpha=1,2,3,4$ taken from the beginning of the Greek
alphabet.\footnote{Since $\alpha$ keeps track of whether $x$ is an upper index
($\alpha=1,3$) or a lower index ($\alpha=2,4$) from now on we need not
distinguish between them: $\rho_{x}=\rho^{x}=\rho(x)$.}

Curves in $T^{\ast}\mathbf{C}$ allow us to define vectors: the vector $\bar
{V}$ tangent to the curve $X=X(\lambda)$ parametrized by $\lambda$ at the
point $X(\lambda)$ is written as
\begin{equation}
\bar{V}=V^{\alpha x}\frac{\delta}{\delta X^{\alpha x}}~,\quad\text{where}\quad
V^{\alpha x}=\frac{dX^{\alpha x}}{d\lambda}~.
\end{equation}
where the repeated indices indicate a summation over $\alpha$ and an
integration over $x$. The directional derivative of a functional $F[X]$ along
the curve $X(\lambda)$ is
\begin{equation}
\frac{dF}{d\lambda}=\mathsf{\tilde{\nabla}}F[\bar{V}]=\frac{\delta F}{\delta
X^{\alpha x}}V^{\alpha x}~, \label{dir deriv}%
\end{equation}
where $\mathsf{\tilde{\nabla}}$ is the functional gradient in $T^{\ast
}\mathbf{C}$,
\begin{equation}
\mathsf{\tilde{\nabla}}F=\,\frac{\delta F}{\delta X^{\alpha x}}\mathsf{\tilde
{\nabla}}X^{\alpha x}~. \label{gradient}%
\end{equation}
The tilde serves to distinguish the functional gradient $\tilde{\nabla}$ on
$T^{\ast}\mathbf{C}$ from the spatial gradient $\nabla f=\partial_{a}f\nabla
x^{a}$ on $\mathbf{R}^{3}$.

The fact that probabilities in (\ref{P}) are normalized introduces a technical
difficulty in that the coordinates $\rho_{x}$ are not independent. We work our
way around this problem by embedding the $\infty$-dimensional manifold
$\mathbf{C}$ in an $(\infty+1)$-dimensional manifold $\mathbf{C}^{+}$ where
the coordinates $\rho_{x}$ are unconstrained.\footnote{It is this embedding
that eventually leads to quantum states being represented by rays (and not
vectors) in a Hilbert space.} Then, since $\mathsf{\tilde{\nabla}}F$ is
strictly speaking a covector on $T^{\ast}\mathbf{C}^{+}$, in order to obtain
the desired directional derivatives (\ref{dir deriv}) on $T^{\ast}\mathbf{C}$
we must impose that the vectors $\bar{V}$ be restricted to be tangent to
$\mathbf{C}$. Such tangent vectors are constrained to obey
\begin{equation}
\frac{d}{d\lambda}\int dx\rho_{x}\,=\int dx\,\frac{dX^{1x}}{d\lambda}=0~.
\label{tang vector}%
\end{equation}

Once local coordinates $(\rho,\Phi,\phi,\rho_{s})$ on $T^{\ast}\mathbf{C}^{+}$
have been introduced there is a natural choice of symplectic form:
\begin{equation}
\Omega=\int dx\,\left[  \tilde{\nabla}\rho^{x}\otimes\tilde{\nabla}\Phi
_{x}-\tilde{\nabla}\Phi_{x}\otimes\tilde{\nabla}\rho^{x}+\tilde{\nabla}%
\phi^{x}\otimes\tilde{\nabla}\rho_{sx}-\tilde{\nabla}\rho_{sx}\otimes
\tilde{\nabla}\phi^{x}\right]  ~. \label{sympl form a}%
\end{equation}
The action of $\Omega$ on two vectors $\bar{V}=d/d\lambda$ and $\bar{U}%
=d/d\mu$ is given by
\begin{equation}
\Omega\lbrack\bar{V},\bar{U}]=\int dx\,\left[  V^{1x}U^{2x}-V^{2x}%
U^{1x}+V^{3x}U^{4x}-V^{4x}U^{3x}\right]  =\Omega_{\alpha x,\beta x^{\prime}%
}V^{\alpha x}U^{\beta x^{\prime}}~,~ \label{sympl form b}%
\end{equation}
and the components of $\Omega$ are
\begin{equation}
\Omega_{\alpha x,\beta x^{\prime}}=%
\begin{bmatrix}
0 & 1 & 0 & 0\\
-1 & 0 & 0 & 0\\
0 & 0 & 0 & 1\\
0 & 0 & -1 & 0
\end{bmatrix}
\,\delta(x,x^{\prime})~. \label{sympl form c}%
\end{equation}

\subsection{Hamiltonian flows and Poisson brackets}

We seek a dynamics that preserves the symplectic structure. Specifically we
are interested in those vector fields $V^{\alpha x}[X]=dX^{\alpha x}/d\lambda$
that generate flows $X^{\alpha x}=X^{\alpha x}(\lambda)$ that preserve the
symplectic form,
\begin{equation}
\pounds _{V}\Omega=0~,
\end{equation}
where the Lie derivative \cite{Schutz 1980} is given by
\begin{equation}
(\pounds _{V}\Omega)_{\alpha x,\beta x^{\prime}}=V^{\gamma x^{\prime\prime}%
}\tilde{\nabla}_{\gamma x^{\prime\prime}}\Omega_{\alpha x,\beta x^{\prime}%
}+\Omega_{\gamma x^{\prime\prime},\beta x^{\prime}}\tilde{\nabla}_{\alpha
x}V^{\gamma x^{\prime\prime}}+\Omega_{\alpha x,\gamma x^{\prime\prime}}%
\tilde{\nabla}_{\beta x^{\prime}}V^{\gamma x^{\prime\prime}}~.
\end{equation}
As discussed in \cite{Schutz 1980}\cite{Arnold 1997} the field $V^{\alpha x}$
is such that the covector $\Omega_{\beta x^{\prime},\alpha x}V^{\beta
x^{\prime}}$ is the gradient of a scalar function, which we will denote
$\tilde{V}[X]$,%
\begin{equation}
\Omega_{\alpha x,\beta x^{\prime}}V^{\beta x^{\prime}}=-\tilde{\nabla}_{\alpha
x}\tilde{V}~. \label{grad V}%
\end{equation}
More explicitly, this condition is written as
\begin{equation}
\frac{d\rho_{x}}{d\lambda}=\frac{\delta\tilde{V}}{\delta\Phi_{x}},\quad
\frac{d\Phi_{x}}{d\lambda}=-\frac{\delta\tilde{V}}{\delta\rho_{x}},\quad
\frac{d\phi_{x}}{d\lambda}=\frac{\delta\tilde{V}}{\delta\rho_{sx}}%
,~\text{and}\quad\frac{d\rho_{sx}}{d\lambda}=-\frac{\delta\tilde{V}}%
{\delta\phi_{x}}, \label{Hamiltonian flow a}%
\end{equation}
which we recognize as Hamilton's equations. The vector $\bar{V}$ is called the
Hamiltonian vector field associated to the Hamiltonian function $\tilde{V}$.

From (\ref{sympl form b}) and (\ref{Hamiltonian flow a}), the action of
$\Omega$ on two Hamiltonian vector fields $\bar{V}=d/d\lambda$ and $\bar
{U}=d/d\mu$ generated respectively by $\tilde{V}$ and $\tilde{U}$ is%
\begin{equation}
\Omega\lbrack\bar{V},\bar{U}]=\int dx\,\left[  \frac{\delta\tilde{V}}%
{\delta\rho_{x}}\frac{\delta\tilde{U}}{\delta\Phi_{x}}-\frac{\delta\tilde{V}%
}{\delta\Phi_{x}}\frac{\delta\tilde{U}}{\delta\rho_{x}}+\frac{\delta\tilde{V}%
}{\delta\phi_{x}}\frac{\delta\tilde{U}}{\delta\rho_{sx}}-\frac{\delta\tilde
{V}}{\delta\rho_{sx}}\frac{\delta\tilde{U}}{\delta\phi_{x}}\right]
\overset{\text{def}}{=}\{\tilde{V},\tilde{U}\}~, \label{PB a}%
\end{equation}
where on the right we introduced the Poisson bracket notation. Thus, the
action of $\Omega$ on two Hamiltonian vector fields is the Poisson bracket of
the associated Hamiltonian functions,%
\begin{equation}
\Omega\lbrack\bar{V},\bar{U}]=\Omega_{\alpha x,\beta x^{\prime}}V^{\alpha
x}U^{\beta x^{\prime}}=\{\tilde{V},\tilde{U}\}~. \label{PB b}%
\end{equation}
The main idea of this section can now be expressed succinctly: the condition
for a flow generated by the vector field $V^{\alpha x}$ to preserve the
symplectic structure, $\pounds _{V}\Omega=0$, is that $V^{\alpha x}$ be the
Hamiltonian vector field associated to a Hamiltonian function $\tilde{V}$,
eq.(\ref{Hamiltonian flow a}),
\begin{equation}
V^{\alpha x}=\frac{dX^{\alpha x}}{d\lambda}=\{X^{\alpha x},\tilde{V}\}~.
\label{Hamiltonian flow b}%
\end{equation}

At this point our goal of defining an ED that preserves the symplectic
structure $\Omega$ and reproduces the continuity equation (\ref{ED c}) is
partially achieved: it is given by the Hamiltonian flow of the scalar
functional $\tilde{H}$ in (\ref{Hamiltonian a}). The remaining task is to
determine the as yet unspecified integration constant $F[\rho,\phi,\theta]$ in
(\ref{Hamiltonian a}).

\subsection{The normalization constraint}

\label{normalization constraint}The conservation of probability implied by the
continuity equation is expressed as a constraint,
\begin{equation}
\tilde{N}=0\quad\text{where}\quad\tilde{N}=1-\left\vert \rho\right\vert
\quad\text{and}\quad\left\vert \rho\right\vert \overset{\text{def}}{=}\int
dx\,\rho(x)~, \label{N constraint}%
\end{equation}
that must be preserved by the Hamiltonian flow,
\begin{equation}
\partial_{t}\tilde{N}=\{\tilde{N},\tilde{H}\}=0~. \label{N conservation}%
\end{equation}
The constraint $\tilde{N}$ itself generates a Hamiltonian flow given by the
vector field
\begin{equation}
\bar{N}=N^{\alpha x}\frac{\delta}{\delta X^{\alpha x}}\quad\text{with}\quad
N^{\alpha x}=\frac{dX^{\alpha x}}{d\sigma}=\{X^{\alpha x},\tilde{N}\}~,
\label{N vector}%
\end{equation}
where $\sigma$ is the parameter along the flow lines. More explicitly,
\begin{equation}
N^{1x}=\frac{d\rho_{x}}{d\sigma}=0,~N^{2x}=\frac{d\Phi_{x}}{d\sigma}%
=1,~N^{3x}=\frac{d\phi_{x}}{d\sigma}=0,~\text{and}~N^{4x}=\frac{d\rho_{sx}%
}{d\sigma}=0. \label{N vector b}%
\end{equation}
The conservation of $\tilde{N}$, eq.(\ref{N conservation}), implies that
$\tilde{N}$ is the generator of a symmetry, namely,
\begin{equation}
\frac{d\tilde{H}}{d\sigma}=\{\tilde{H},\tilde{N}\}=0~.
\end{equation}
Integrating (\ref{N vector b}) one finds the integral curves generated by
$\tilde{N}$,
\begin{equation}
\rho_{x}(\sigma)=\rho_{x}(0)\,,~\Phi_{x}(\sigma)=\Phi_{x}(0)+\sigma
\,,~\phi_{x}(\sigma)=\phi_{x}(0)\,,~\text{and}~\rho_{sx}(\sigma)=\rho
_{sx}(0)~, \label{ray eq}%
\end{equation}
which shows that the symmetry generated by $\tilde{N}$ is to shift the phase
$\Phi$ by a constant $\sigma$ without otherwise changing the dynamics. The
interpretation is that the constraint $\tilde{N}=0$ reduces by one the
(infinite) number of independent $\rho_{x}$ degrees of freedom and it also
reduces by one the number of $\Phi_{x}$s because for any value of $\sigma
$\ the phases $\Phi_{x}+\sigma$\ and $\Phi_{x}$\ correspond to the same state.
The result is a global gauge symmetry. (This is the ED analogue of the fact
that in QM states are represented by rays rather than vectors in a Hilbert space.)

The situation thus far can be summarized as follows: the phase space of
interest is $T^{\ast}\mathbf{C}$ but the constraint $|\rho|{}=1$ forces us to
use coordinates in a larger embedding space $T^{\ast}\mathbf{C}^{+}$. The
introduction of one superfluous coordinate has led us to introduce the
corresponding superfluous momentum. We eliminate the extra coordinate by
imposing $\tilde{N}=0$ and we eliminate the extra momentum by declaring it an
unphysical gauge variable.

\section{The information geometry of e-phase space}

We have just seen that a natural criterion to update the constraints is to
impose a dynamics that preserves a symplectic structure. To select the
particular Hamiltonian dynamics that reproduces the QM of spin we proceed in
close analogy to the ED of scalar particles. The procedure is to generalize
the metric structure that is naturally available in the statistical manifold
$\mathbf{P}$ $=\{\rho\}$ --- its information geometry --- to the full phase
space $T^{\ast}\mathbf{C}=\{\rho,\Phi,\phi,\rho_{s}\}$. Then the criterion we
adopt for updating constraints is a dynamics that preserves both the
symplectic and the metric structures.

In this section the objective is to transform e-phase space $T^{\ast
}\mathbf{C}$ from a manifold that is merely symplectic to a manifold that is
both symplectic and Riemannian. Then we will find the corresponding
Hamilton-Killling flows (section \ref{HK flows}) and the particular
Hamiltonians (section \ref{H and P eq}) that reproduce the ED of spin,
eqs.(\ref{ED c}, \ref{Hamiltonian flow a}).

\subsection{The metric on the embedding space $T^{\ast}\mathbf{C}^{+}$}

\label{metric in T*C+1}As we saw earlier the normalization constraint forces
us to embed the phase space of interest $T^{\ast}\mathbf{C}$ in a larger
unconstrained space $T^{\ast}\mathbf{C}^{+}$. The first goal is to extend the
information geometry of the statistical manifold $\mathbf{P}$ $=\{\rho\}$ to
the full e-phase space $T^{\ast}\mathbf{C}^{+}=\{\rho,\Phi,\phi,\rho_{s}\}$.
In the next subsection we obtain the metric induced on the space $T^{\ast
}\mathbf{C}$.

One possible path (pursued in \cite{Carrara 2021}) is to proceed exactly as in
the case of scalar particles \cite{Caticha 2019} where the known spherical
symmetry of the statistical manifold $\mathbf{P}$ $=\{\rho\}$ is extended to
the unconstrained embedding space $T^{\ast}\mathbf{P}^{+}$. It is very
significant that the geometry of the embedding space $T^{\ast}\mathbf{P}^{+}$
is not unique; any spherically symmetric geometry will serve our purposes. We
can therefore choose the simplest possible embedding in which the space
$T^{\ast}\mathbf{P}^{+}$ happens to be a flat. This choice, while strictly
optional, turns out to be very convenient for calculational purposes because
it allows one to introduce the notion of Hilbert spaces and leads to a linear
Schr\"{o}dinger equation.

Here we follow an alternative path: rather than extending the metric from
$\mathbf{P}$ to $T^{\ast}\mathbf{C}^{+}$, we start from the already known
metric for $T^{\ast}\mathbf{P}^{+}$ derived for the scalar case in
\cite{Caticha 2019} and extend it to $T^{\ast}\mathbf{C}^{+}$. Furthermore,
from the beginning we choose both spaces $T^{\ast}\mathbf{P}^{+}$ and
$T^{\ast}\mathbf{C}^{+}$ to be flat. This approach is designed to guarantee
that the resulting metric will have the desired spherical symmetry and that it
will correctly reproduce the scalar limit. The argument is simplest when we
adopt complex coordinates. In the scalar case the metric of the flat space
$T^{\ast}\mathbf{P}^{+}$ is given by\footnote{See eq.(91) in \cite{Caticha
2019}. The choice $A=0$ and $B=1/2\hbar$ yields a flat geometry. The first
term $(\delta\rho)^{2}/\rho$ in (\ref{scalar metric}) is the information
metric on the base space $\mathbf{P}^{+}$; the second term $\rho(\delta
\varphi)^{2}$ is the metric on the cotangent fibers of the bundle $T^{\ast
}\mathbf{P}^{+}$.}
\begin{equation}
\delta\ell_{0}^{2}=\frac{1}{4}\int dx\,\left[  \frac{1}{\rho_{x}}(\delta
\rho_{x})^{2}+\rho_{x}(\delta\varphi)^{2}\right]  =\int dx\,\delta\psi
_{x}^{\ast}\delta\psi_{x}~, \label{scalar metric}%
\end{equation}
where
\begin{equation}
\psi=\rho^{1/2}e^{i\varphi}~ \label{scalar psi}%
\end{equation}
is the scalar wave function, $\varphi$ is the drift potential, and
$\hbar\varphi$ is the momentum conjugate to $\rho$.

\paragraph*{Introducing the spinor wave function ---}

In Geometric Algebra spinors are introduced as elements of a minimal left
ideal \cite{Riesz 1958} \cite{Hestenes 1966} (see Appendix A.2). We propose
that the natural generalization of the scalar wave function (\ref{scalar psi})
that takes the peculiar rotational properties of spin into account is given by
the spinor wave function
\begin{equation}
\Psi=\rho^{1/2}e^{i\varphi}U(\theta,\phi,\chi)u_{+}~, \label{spinor psi a}%
\end{equation}
where the basis spinors that represent spin up and down along the $\vec{e}%
_{3}$ direction are
\begin{equation}
u_{+}=\frac{1}{\sqrt{2}}\left(  1+\vec{e}_{3}\right)  \quad\text{and}\quad
u_{-}=\vec{e}_{1}u_{+}~,
\end{equation}
that is, $\vec{e}_{3}u_{\pm}=\pm u_{\pm}$. The rotor $U(\theta,\phi,\chi)$
acts on the basis spinor $u_{+}$ to yield a generic normalized spinor $u$
given by eq.(\ref{spinor d}),
\begin{equation}
u=U(\theta,\phi,\chi)u_{+}=u_{+}e^{-i(\chi+\phi)/2}\cos\frac{\theta}{2}%
+u_{-}e^{-i(\chi-\phi)/2}\sin\frac{\theta}{2}~. \label{spinor d a}%
\end{equation}

We can now address the question about the still unspecified value of the
multiplier $\gamma$ in eq.(\ref{phase a}). First we use eq.(\ref{U Euler a})
to write $\Psi$ in the alternative forms
\begin{equation}
\Psi=\rho^{1/2}U(\theta,\phi,\bar{\chi})u_{+}=\rho^{1/2}e^{-i\bar{\chi}%
/2}U(\theta,\phi,0)u_{+}\quad\text{where}\quad\bar{\chi}=\chi-2\varphi~,
\label{spinor psi b}%
\end{equation}
which show that the phase of $\Psi$\ can be interpreted as a rotation of the
spin frame by an angle $\bar{\chi}$ about the spin vector $\vec{s}$. Then we
argue that just as the phase $\Phi$ of the scalar wave function, $\psi
=\rho^{1/2}e^{i\Phi/\hbar}$ in eq.(\ref{scalar psi}), is the momentum
conjugate to $\rho$ \cite{Caticha 2019}, in the spin case we are also led to
identify the momentum conjugate to $\rho$ in (\ref{ED c}) with the phase of
the spinor wave function $\Psi$ in eq.(\ref{spinor psi b}), that is,
\begin{equation}
\Psi=\rho^{1/2}e^{i\Phi/\hbar}U(\theta,\phi,0)u_{+}~. \label{spinor psi c}%
\end{equation}
Therefore, in order to reproduce the rotational properties of spin $1/2$ the
multiplier $\gamma$ in eq.(\ref{phase a}) must be chosen to be $\gamma=1/2$,
so that the phase is
\begin{equation}
\Phi=\hbar\left(  \varphi-\frac{1}{2}\chi\right)  =-\frac{\hbar}{2}\bar{\chi
}~. \label{phase b}%
\end{equation}
Furthermore, since
\begin{equation}
\Psi\Psi^{\dagger}=\rho Uu_{+}u_{+}^{\dagger}U^{\dagger}=\rho U(1+\vec{e}%
_{3})U^{\dagger}=\rho(1+\vec{s})~,
\end{equation}
we have an elegant extension of the Born rule in which both the probability
density and the spin density can be written in terms of $\Psi\Psi^{\dagger}$:
\begin{equation}
\langle\Psi\Psi^{\dagger}\rangle_{0}=\rho\quad\text{and}\quad\langle\Psi
\Psi^{\dagger}\rangle_{1}=\rho\vec{s}~. \label{Born rule}%
\end{equation}

\paragraph*{Complex coordinates ---}

A useful new set of coordinates is suggested by using eq.(\ref{spinor d a})
which leads us to write (\ref{spinor psi b}) as
\begin{equation}
\Psi=\psi_{+}u_{+}+\psi_{-}u_{-}~, \label{spinor psi d}%
\end{equation}
where
\begin{equation}
\psi_{+}=\rho^{1/2}\cos\frac{\theta}{2}e^{-i(\bar{\chi}+\phi)/2}%
\quad\text{and}\quad\psi_{-}=\rho^{1/2}\sin\frac{\theta}{2}e^{-i(\bar{\chi
}-\phi)/2}~ \label{psi +-}%
\end{equation}
are called the complex \textquotedblleft amplitudes\textquotedblright\ for
spin up and down respectively. A straightforward calculation of the symplectic
form $\Omega$, eq.(\ref{sympl form a}), in terms of $\psi_{\pm}$, shows that
the coordinate transformation
\begin{equation}
\left(  \rho,\Phi,\phi,\rho_{s}\right)  \rightarrow(\psi_{+},i\mathsf{\hbar
}\psi_{+}^{\ast},\psi_{-},i\mathsf{\hbar}\psi_{-}^{\ast})~,
\end{equation}
is indeed canonical. The new coordinates of a point $\Psi$,
\begin{equation}
\Psi^{\mu x}=(\Psi^{1x},\Psi^{2x},\Psi^{3x},\Psi^{4x})=(\psi_{+x}%
,i\mathsf{\hbar}\psi_{+x}^{\ast},\psi_{-x},i\mathsf{\hbar}\psi_{-x}^{\ast})~,
\label{spinor psi e}%
\end{equation}
will be labeled by a composite index $\mu x$ with the Greek index $\mu$ taken
from the middle of the Greek alphabet. In these new coordinates the action of
$\Omega$ on two vectors $\bar{V}=d/d\lambda$ and $\bar{U}=d/d\mu$ is given by
\begin{equation}
\Omega\lbrack\bar{V},\bar{U}]=\Omega_{\mu x,\nu x^{\prime}}V^{\mu x}U^{\nu
x^{\prime}}~,~ \label{sympl form d}%
\end{equation}
where the components of $\Omega$ are
\begin{equation}
\Omega_{\mu x,\nu x^{\prime}}=%
\begin{bmatrix}
0 & 1 & 0 & 0\\
-1 & 0 & 0 & 0\\
0 & 0 & 0 & 1\\
0 & 0 & -1 & 0
\end{bmatrix}
\,\delta(x,x^{\prime})~. \label{sympl form e}%
\end{equation}

\paragraph*{The metric on $T^{\ast}\mathbf{C}^{+}$ ---}

We propose that the natural generalization of the scalar metric
(\ref{scalar metric}) from the space $T^{\ast}\mathbf{P}^{+}$ to the space
$T^{\ast}\mathbf{C}^{+}$ is
\begin{equation}
\delta\ell^{2}=\int dx\langle\,\delta\Psi_{x}^{\dagger}\delta\Psi_{x}%
\rangle_{0}~. \label{spinor metric a}%
\end{equation}
To get insight recall eq.(\ref{spinor psi b}) and introduce the quaternion
variable
\begin{equation}
\Upsilon=\rho^{1/2}U(\theta,\phi,\bar{\chi})~.
\end{equation}
Then
\begin{equation}
\delta\ell^{2}=\int dx\langle\,u_{+}^{\dagger}\delta\Upsilon^{\dagger}%
\delta\Upsilon u_{+}\rangle_{0}=\int dx\langle\,(1+\vec{e}_{3})\delta
\Upsilon^{\dagger}\delta\Upsilon\rangle_{0}~.
\end{equation}
Next use (\ref{dU zeta}) and (\ref{dzeta a}) to write
\begin{align}
\delta\Upsilon^{\dagger}\delta\Upsilon &  =\delta(\rho^{1/2}U^{\dagger}%
)\delta(\rho^{1/2}U)=\delta(\rho^{1/2})^{2}+\rho\delta U^{\dagger}\delta
U\nonumber\\
&  =\frac{1}{4\rho_{x}}(\delta\rho_{x})^{2}+\frac{1}{4}\rho_{x}(\delta
\vec{\zeta})^{2}~,
\end{align}
where
\begin{equation}
\delta\vec{\zeta}=\vec{e}_{3}\delta\phi+\vec{e}_{\phi}\delta\theta+\vec
{s}\,\delta\bar{\chi}~. \label{dzeta b}%
\end{equation}
Since $\delta\Upsilon^{\dagger}\delta\Upsilon$ is a scalar we have
\begin{equation}
\delta\ell^{2}=\int dx\,\delta\Upsilon^{\dagger}\delta\Upsilon=\frac{1}{4}\int
dx\left(  \frac{1}{\rho_{x}}(\delta\rho_{x})^{2}+\rho_{x}(\delta\vec{\zeta
})^{2}\right)  ~, \label{spinor metric b}%
\end{equation}
which shows that the angle $\delta\varphi$ in the scalar metric,
eq.(\ref{scalar metric}), is replaced by the rotation angle $|\delta\vec
{\zeta}|$. In terms of the complex coordinates $\psi_{\pm}$ the metric
(\ref{spinor metric a}) takes a particularly simple form,
\begin{equation}
\delta\ell^{2}=\int dx\left(  \,\delta\psi_{+x}^{\ast}\delta\psi_{+x}%
+\delta\psi_{-x}^{\ast}\,\delta\psi_{-x}\right)  ~. \label{spinor metric c}%
\end{equation}
The scalar product of two vectors $\bar{V}=d/d\lambda$ and $\bar{U}=d/d\mu$ is
given by
\begin{equation}
G[\bar{V},\bar{U}]=G_{\mu x,\nu x^{\prime}}V^{\mu x}U^{\nu x^{\prime}}~,~
\label{spinor metric d}%
\end{equation}
where the components of $G^{+}$ and its inverse are
\begin{equation}
G_{\mu x,\nu x^{\prime}}=\frac{1}{2i\hbar}%
\begin{bmatrix}
0 & 1 & 0 & 0\\
1 & 0 & 0 & 0\\
0 & 0 & 0 & 1\\
0 & 0 & 1 & 0
\end{bmatrix}
\,\delta(x,x^{\prime})~,\quad G^{\mu x,\nu x^{\prime}}=2i\hbar%
\begin{bmatrix}
0 & 1 & 0 & 0\\
1 & 0 & 0 & 0\\
0 & 0 & 0 & 1\\
0 & 0 & 1 & 0
\end{bmatrix}
\,\delta(x,x^{\prime}). \label{spinor metric e}%
\end{equation}

Yet another important aspect of the metric (\ref{spinor metric a}) can be made
explicit in the following coordinates,
\begin{equation}
\psi_{+}=\xi_{1}+i\xi_{2}\quad\text{and}\quad\psi_{-}=\xi_{3}+i\xi_{4}~.
\end{equation}
Then the normalization condition,
\begin{equation}
\int dx\left(  \,\psi_{+x}^{\ast}\psi_{+x}+\psi_{-x}^{\ast}\,\psi_{-x}\right)
=\int dx\,\rho_{x}=1~,
\end{equation}
becomes
\begin{equation}
\int dx\,\left(  \xi_{1x}^{2}+\xi_{2x}^{2}+\xi_{3x}^{2}+\xi_{4x}^{2}\right)
=1~,
\end{equation}
which is the equation of the unit sphere in $T^{\ast}\mathbf{C}^{+}$. This
shows that our choice (\ref{spinor metric a}) which extends the information
geometry of the statistical manifold $\mathbf{P}$ $=\{\rho\}$ to the space
$T^{\ast}\mathbf{C}^{+}=\{\rho,\Phi,\phi,\rho_{s}\}$ amounts to extending the
well-known infinite dimensional spherical symmetry of the space $\mathbf{P}$
to the larger embedding space $T^{\ast}\mathbf{C}^{+}$. Furthermore, the
metric (\ref{spinor metric c}) becomes
\begin{equation}
\delta\ell^{2}=\int dx\left(  \,\delta\xi_{1x}^{2}+\delta\xi_{2x}^{2}%
+\delta\xi_{3x}^{2}+\delta\xi_{4x}^{2}\right)  ~,
\end{equation}
which is Euclidean. This means that, beyond being spherically symmetric, the
particular embedding space $T^{\ast}\mathbf{C}^{+}$ that we have chosen is flat.

\paragraph*{A complex structure on $T^{\ast}\mathbf{C}^{+}$ ---}

The contraction of the symplectic form $\Omega$, eq.(\ref{sympl form d}), with
the inverse metric $G^{-1}$ allows us to construct a mixed tensor $J$ with
components
\begin{equation}
J^{\mu x}{}_{\nu x^{\prime}}=-\frac{1}{2\hbar}\,G^{\mu x,\lambda
x^{\prime\prime}}\Omega_{\lambda x^{\prime\prime},\nu x^{\prime}}=%
\begin{bmatrix}
i & 0 & 0 & 0\\
0 & -i & 0 & 0\\
0 & 0 & i & 0\\
0 & 0 & 0 & -i
\end{bmatrix}
\,\delta(x,x^{\prime})~. \label{J tensor}%
\end{equation}
What makes the tensor $J$ special is that its square is
\begin{equation}
J^{\mu x}{}_{\lambda x^{\prime\prime}}J^{\lambda x^{\prime\prime}}{}_{\nu
x^{\prime}}=-\delta^{\mu x}{}_{\nu x^{\prime}}~.
\end{equation}
In words, the action of $J^{2}$ (or $\Omega^{2}$) is equivalent to multiplying
by $-1$. This means that $J$ plays the role of a complex structure.

Note, however, that the $i$ that appears in all these expressions, including
in the complex structure $J$, is not the imaginary unit familiar from complex
analysis but the pseudoscalar $i$, eq.(\ref{pseudoscalar}), that is
geometrically interpreted as a directed volume. Strictly speaking we are not
using complex variables.

\subsection{The metric induced on $T^{\ast}\mathbf{C}$}

As we saw in section \ref{normalization constraint} the e-phase space
$T^{\ast}\mathbf{C}$ is obtained from the embedding space $T^{\ast}%
\mathbf{C}^{+}$ by the restriction $|\rho|\,=1$ and by identifying the point
$X=(\rho_{x},\Phi_{x},\phi_{x},\rho_{sx})$ with the shifted points $X_{\sigma
}=(\rho_{x},\Phi_{x}+\sigma,\phi_{x},\rho_{sx})$ that lie on the same ray.

Consider two neighboring\ points $X$ and $X^{\prime}=X+\delta X$. The metric
induced on $T^{\ast}\mathbf{C}$ is defined as the shortest $T^{\ast}%
\mathbf{C}^{+}$ distance between $X$ and points $X_{\sigma}^{\prime}$ on the
ray defined by $X^{\prime}$. From (\ref{phase b}) the effect of shifting
$\Phi_{x}$ by $\sigma$ is to shift $\bar{\chi}$ by $\bar{\sigma}%
=-2\sigma/\hbar$. Then, setting $|\rho|\,=1$ and recalling (\ref{dzeta b}) and
(\ref{spinor metric b}), the $T^{\ast}\mathbf{C}^{+}$ distance between $X$ and
$X_{\sigma}^{\prime}$ is given by%
\begin{equation}
\delta\ell_{\sigma}^{2}=\frac{1}{4}\int dx\left(  \frac{1}{\rho}(\delta
\rho)^{2}+\rho(\delta\vec{\zeta}+\vec{s}\,\delta\bar{\sigma})^{2}\right)  ~,
\label{TC metric a}%
\end{equation}
Let%
\begin{equation}
\delta\ell_{FS}^{2}=\min_{\delta\bar{\sigma}}\delta\ell_{\sigma}^{2}~.
\label{TC metric b}%
\end{equation}
The value of $\delta\bar{\sigma}$ that minimizes (\ref{TC metric a}) is
\begin{equation}
\delta\bar{\sigma}=-\int dx\rho\,\vec{s}\cdot\delta\vec{\zeta}=-\,\langle
\vec{s}\cdot\delta\vec{\zeta}\rangle
\end{equation}
so that the metric on $T^{\ast}\mathbf{C}$ is
\begin{equation}
\delta\ell_{FS}^{2}=\frac{1}{4}\int dx\left[  \frac{1}{\rho}(\delta\rho
)^{2}+\rho\left(  \delta\vec{\zeta}-\vec{s}\,\langle\vec{s}\cdot\delta
\vec{\zeta}\rangle\right)  ^{2}\right]  ~, \label{TC metric c}%
\end{equation}
which can be recognized as the Fubini-Study metric for a spin-$1/2$ particle.

\paragraph*{A convenient choice of gauge ---}

The scalar product between two vectors $\bar{V}=d/d\lambda$ and $\bar
{U}=d/d\mu$ is a bit messy,
\begin{equation}
G(\bar{V},\bar{U})=\frac{1}{4}\int dx\,\left[  \frac{1}{\rho}\frac{d\rho
}{d\lambda}\frac{d\rho}{d\mu}+\rho_{x}\left(  \frac{d\vec{\zeta}}{d\lambda
}-\vec{s}\,\langle\vec{s}\cdot\frac{d\vec{\zeta}}{d\lambda}\rangle\right)
\left(  \frac{d\vec{\zeta}}{d\mu}-\vec{s}\,\langle\vec{s}\cdot\frac
{d\vec{\zeta}}{d\mu}\rangle\right)  \right]  ~, \label{TC metric d}%
\end{equation}
but can be simplified considerably. First, we note that by virtue of being
tangent to $T^{\ast}\mathbf{C}$ a vector $\bar{V}$ satisfies
\begin{equation}
|\frac{d\rho}{d\lambda}|{}=\int dx\,\frac{d\rho_{x}}{d\lambda}=0~.
\label{TGF a}%
\end{equation}
Furthermore, since points $X$ on the same ray are equivalent it follows all
vectors $\bar{V}$ and $\bar{V}^{\prime}$ that differ by a vector along the
gauge direction $\bar{N}$ are equivalent. To prove this consider a curve
$X(\lambda)$ and a second curve $X^{\prime}(\lambda)$ obtained from the first
by a shift along the gauge direction. We assume that their coordinates are
identical except for
\begin{equation}
\Phi_{x}^{\prime}(\lambda)=\Phi_{x}(\lambda)+k(\lambda-\lambda_{0})
\end{equation}
so that the two curves are equivalent and they cross at $\lambda=\lambda_{0}$.
Using eq.(\ref{N vector b}) we see that the tangent vectors at $\lambda_{0}$
are related by
\begin{equation}
\frac{dX^{\prime\alpha x}}{d\lambda}=\frac{dX^{\alpha x}}{d\lambda}+kN^{\alpha
x}\quad\text{or}\quad\bar{V}^{\prime}=\bar{V}+k\bar{N}~.
\end{equation}
To work with such equivalent vectors it desirable to choose a convenient
representative, that is, we fix the gauge. Equation (\ref{TC metric d})
suggests that a convenient \textquotedblleft Tangent
Gauge-Fixed\textquotedblright\ representative (which we will call the TGF
gauge) is
\begin{equation}
\int dx\rho\,\vec{s}\cdot\frac{d\vec{\zeta}}{d\lambda}=\langle\vec{s}%
\cdot\frac{d\vec{\zeta}}{d\lambda}\rangle=0~. \label{TGF b}%
\end{equation}
Vectors satisfying (\ref{TGF a}) and (\ref{TGF b}) will be called TGF vectors:
the first condition enforces a flow tangent to the $\left\vert \rho\right\vert
=1$ surface; the second condition eliminates a superfluous vector component
along the gauge direction $\bar{N}$. In the TGF gauge $G(\bar{V},\bar{U})$
simplifies to
\begin{equation}
G(\bar{V},\bar{U})=\frac{1}{4}\int dx\,\left[  \frac{1}{\rho}\frac{d\rho
}{d\lambda}\frac{d\rho}{d\mu}+\rho\frac{d\vec{\zeta}}{d\lambda}\cdot
\frac{d\vec{\zeta}}{d\mu}\right]  ~. \label{TC metric e}%
\end{equation}
To be explicit: the tensor $G$ in (\ref{TC metric e}) can act on arbitrary
vectors but it is only when they satisfy the TGF conditions (\ref{TGF a}) and
(\ref{TGF b}) that $G$ can be interpreted as the metric on $T^{\ast}%
\mathbf{C}$. The corresponding expression for the length element is
\begin{equation}
\delta\ell_{FS}^{2}=\frac{1}{4}\int dx\left[  \frac{1}{\rho}\delta\rho
^{2}+\rho\delta\vec{\zeta}^{2}\right]  ~, \label{TC metric f}%
\end{equation}
where it is understood that $\delta X$ satisfies the TGF conditions,
\begin{equation}
|\delta\rho|{}=0\quad\text{and}\quad\langle\vec{s}\cdot\delta\vec{\zeta
}\rangle=0~. \label{TGF c}%
\end{equation}

\paragraph*{The $T^{\ast}\mathbf{C}$ metric in $\psi_{\pm}$ coordinates ---}

The same analysis can be carried out in $\psi_{\pm}$ coordinates. The
normalization constraint now reads
\begin{equation}
\tilde{N}=0\quad\text{with }\tilde{N}=1-\int dx\left(  \,\psi_{+}\psi
_{+}^{\ast}+\,\psi_{-}\psi_{-}^{\ast}\right)  ~.
\end{equation}
From eqs.(\ref{ray eq}), (\ref{phase b}) and (\ref{psi +-}) we see that the
Hamiltonian flow generated by $\tilde{N}$ and parametrized by $\sigma$ yields
the integral curves
\begin{equation}
\psi_{\pm}(\sigma)=\psi_{\pm}(0)e^{i\sigma/\hbar}~. \label{ray eq b}%
\end{equation}
Using
\begin{equation}
\left\langle \Psi\frac{d\Psi^{\dagger}}{d\lambda}\right\rangle _{0+3}=\frac
{1}{2}\frac{d\rho}{d\lambda}+i\rho\,\vec{s}\cdot\frac{d\vec{\zeta}}{d\lambda}%
\end{equation}
the two real TGF conditions, (\ref{TGF a}) and (\ref{TGF b}), are replaced by
a single complex condition,
\begin{equation}
\int dx\left\langle \Psi\frac{d\Psi^{\dagger}}{d\lambda}\right\rangle
_{0+3}=\int dx\left(  \,\psi_{+}\frac{d\psi_{+}^{\ast}}{d\lambda}+\,\psi
_{-}\frac{d\psi_{-}^{\ast}}{d\lambda}\right)  =0~.
\end{equation}
In $\psi_{\pm}$ coordinates the metric on $T^{\ast}\mathbf{C}$,
eq.(\ref{TC metric f}), reads%
\begin{equation}
\delta\ell_{FS}^{2}=\int dx\left(  \delta\,\psi_{+}\delta\psi_{+}^{\ast
}+\delta\,\psi_{-}\delta\psi_{-}^{\ast}\right)  =\int dxdx^{\prime}G_{\mu
x,\nu x^{\prime}}\,\delta\Psi^{\mu x}\delta\Psi^{\nu x^{\prime}}~,
\label{TC metric g}%
\end{equation}
where the metric tensor is given by (\ref{spinor metric e}). In terms of the
spinor wave function $\Psi$ the metric on $T^{\ast}\mathbf{C}$ takes the form
(\ref{spinor metric a}),
\begin{equation}
\delta\ell_{FS}^{2}=\int dx\langle\delta\Psi_{x}\delta\Psi_{x}^{\dagger
}\rangle_{0}~.
\end{equation}

\paragraph*{The complex structure on $T^{\ast}\mathbf{C}$ ---}

It is straightforward to check that the tensor $J^{\mu x}{}_{\nu x^{\prime}}$
defined in (\ref{J tensor}) takes a TGF vector $\bar{V}$ to another vector
$J\bar{V}$ that is also TGF. This means that $J^{\mu x}{}_{\nu x^{\prime}}$ is
indeed a tensor on $T^{\ast}\mathbf{C}$ --- it linearly maps vectors to
vectors. Since $J^{2}=-1$ we conclude that the same $J$ that defines a complex
structure on $T^{\ast}\mathbf{C}^{+}$ also serves to define a complex
structure on $T^{\ast}\mathbf{C}$.

\section{Hamilton-Killing flows}

\label{HK flows}Our next goal is to find those Hamiltonian flows $Q^{\mu x}$
that also happen to preserve the metric tensor, that is, $\pounds _{Q}G=0$.
The condition for $Q^{\mu x}$ to be a Killing vector is \cite{Schutz 1980}%
\begin{equation}
(\pounds _{Q}G)_{\mu x,\nu x^{\prime}}=Q^{\lambda x^{\prime\prime}}%
\tilde{\nabla}_{\lambda x^{\prime\prime}}G_{\mu x,\nu x^{\prime}}+G_{\lambda
x^{\prime\prime},\nu x^{\prime}}\tilde{\nabla}_{\mu x}Q^{\lambda
x^{\prime\prime}}+G_{\mu x,\lambda x^{\prime\prime}}\tilde{\nabla}_{\nu
x^{\prime}}Q^{\lambda x^{\prime\prime}}=0~.
\end{equation}
In complex coordinates eqs.(\ref{TC metric g}) and (\ref{spinor metric e})
gives $\tilde{\nabla}_{\gamma x^{\prime\prime}}G_{\alpha x,\beta x^{\prime}%
}=0$, and the Killing equation simplifies to
\begin{equation}
(\pounds _{Q}G)_{\mu x,\nu x^{\prime}}=G_{\lambda x^{\prime\prime},\nu
x^{\prime}}\tilde{\nabla}_{\mu x}Q^{\lambda x^{\prime\prime}}+G_{\mu x,\lambda
x^{\prime\prime}}\tilde{\nabla}_{\nu x^{\prime}}Q^{\lambda x^{\prime\prime}%
}=0~,
\end{equation}
which can be expressed as an $4\times4$ matrix equation
\begin{equation}
(\pounds _{Q}G)_{\mu x,\nu x^{\prime}}=%
\begin{bmatrix}
(\pounds _{Q}G)_{+x,+x^{\prime}} & 0\\
0 & (\pounds _{Q}G)_{-x,-x^{\prime}}%
\end{bmatrix}
=0
\end{equation}
where the $2\times2$ blocks $(\pounds _{Q}G)_{+x,+x^{\prime}}$ and
$(\pounds _{Q}G)_{-x,-x^{\prime}}$ are
\begin{equation}
(\pounds _{Q}G)_{+x,+x^{\prime}}=\frac{1}{2i\hbar}%
\begin{bmatrix}
\frac{\delta Q^{2x^{\prime}}}{\delta\psi_{+x}}+\frac{\delta Q^{2x}}{\delta
\psi_{+x^{\prime}}}~; & \frac{\delta Q^{1x^{\prime}}}{\delta\psi_{+x}}%
+\frac{\delta Q^{2x}}{\delta i\hbar\psi_{+x^{\prime}}^{\ast}}\\
\frac{\delta Q^{2x^{\prime}}}{\delta i\hbar\psi_{+x}^{\ast}}+\frac{\delta
Q^{1x}}{\delta\psi_{+x^{\prime}}}~; & \frac{\delta Q^{1x^{\prime}}}{\delta
i\hbar\psi_{+x}^{\ast}}+\frac{\delta Q^{1x}}{\delta i\hbar\psi_{+x^{\prime}%
}^{\ast}}%
\end{bmatrix}
=0~ \label{K flow a}%
\end{equation}
and
\begin{equation}
(\pounds _{Q}G)_{-x,-x^{\prime}}=\frac{1}{2i\hbar}%
\begin{bmatrix}
\frac{\delta Q^{4x^{\prime}}}{\delta\psi_{-x}}+\frac{\delta Q^{4x}}{\delta
\psi_{-x^{\prime}}}~; & \frac{\delta Q^{3x^{\prime}}}{\delta\psi_{-x}}%
+\frac{\delta Q^{4x}}{\delta i\hbar\psi_{-x^{\prime}}^{\ast}}\\
\frac{\delta Q^{4x^{\prime}}}{\delta i\hbar\psi_{-x}^{\ast}}+\frac{\delta
Q^{3x}}{\delta\psi_{-x^{\prime}}}~; & \frac{\delta Q^{3x^{\prime}}}{\delta
i\hbar\psi_{-x}^{\ast}}+\frac{\delta Q^{3x}}{\delta i\hbar\psi_{-x^{\prime}%
}^{\ast}}%
\end{bmatrix}
=0~. \label{K flow b}%
\end{equation}
But $Q^{\mu x}$ must also generate a Hamiltonian flow, $\pounds _{Q}\Omega=0$.
Substituting
\begin{equation}
Q^{1x}=\frac{\delta\tilde{Q}}{\delta i\hbar\psi_{+x}^{\ast}}\,,~Q^{2x}%
=-\frac{\delta\tilde{Q}}{\delta\psi_{+x}}\,,~Q^{3x}=\frac{\delta\tilde{Q}%
}{\delta i\hbar\psi_{-x}^{\ast}}~,~\text{and}~Q^{4x}=-\frac{\delta\tilde{Q}%
}{\delta\psi_{-x}}~ \label{H flow a}%
\end{equation}
into (\ref{K flow a}) and (\ref{K flow b}) gives us the condition for a
Hamilton-Killing (or HK) flow,
\begin{equation}
\frac{\delta^{2}\tilde{Q}}{\delta\psi_{+x}\delta\psi_{+x^{\prime}}}%
=0,~\frac{\delta^{2}\tilde{Q}}{\delta\psi_{-x}\delta\psi_{-x^{\prime}}%
}=0,~\frac{\delta^{2}\tilde{Q}}{\delta\psi_{+x}^{\ast}\delta\psi_{+x^{\prime}%
}^{\ast}}=0~,~~\text{and~~}\frac{\delta^{2}\tilde{Q}}{\delta\psi_{-x}^{\ast
}\delta\psi_{-x^{\prime}}^{\ast}}=0~. \label{HK flow}%
\end{equation}
The conclusion is that a flow that preserves both $G$ and $\Omega$ requires a
functional $\tilde{Q}[\psi_{+x},\psi_{+x}^{\ast},\psi_{-x},\psi_{-x}^{\ast}]$
that is at most \emph{linear} in each of $\psi$, $\psi_{+}^{\ast}$, $\psi_{-}%
$, and $\psi_{-}^{\ast}$.

The functional form of $\tilde{Q}$ can further be restricted as follows.
First, in order to reflect the rotational properties of spin we shall require
that the various $\psi_{+}$s and $\psi_{-}$s that appear in $\tilde{Q}$ must
combine into spinor wave functions, that is, $\tilde{Q}=\tilde{Q}[\Psi
,\Psi^{\dagger}]$ with $\Psi=\psi_{+}u_{+}+\psi_{-}u_{-}$. And second, we
shall therefore require that $\tilde{Q}[\Psi,\Psi^{\dagger}]$ be invariant
under global gauge transformations,%

\begin{equation}
\tilde{Q}[\Psi e^{i\sigma/\hbar},\Psi^{\dagger}e^{-i\sigma/\hbar}]=\tilde
{Q}[\Psi,\Psi^{\dagger}]~. \label{GGT}%
\end{equation}
The idea is that the HK flows be such that initial states that are equivalent
--- states that lie on the same ray --- flow to final states that are also
equivalent: if the state $\Psi_{x}(\lambda_{i})$ flows to $\Psi_{x}%
(\lambda_{f})$ then $e^{i\sigma/\hbar}\Psi_{x}(\lambda_{i})$ must flow to
$e^{i\sigma/\hbar}\Psi_{x}(\lambda_{f})$. As shown in Appendix B these
conditions constrain $\tilde{Q}$ to be linear in $\Psi$ and in its adjoint
$\Psi^{\dagger}$,
\begin{equation}
\tilde{Q}[\Psi,\Psi^{\dagger}]=\int dx_{1}dx_{2}\langle\Psi_{x_{1}}^{\dagger
}\hat{Q}_{x_{1}x_{2}}\Psi_{x_{2}}\rangle_{0}\quad\text{where }\quad\hat
{Q}_{x_{2}x_{1}}^{\dagger}=\hat{Q}_{x_{1}x_{2}}\, \label{Q funct}%
\end{equation}
is a Hermitian kernel.

\paragraph*{To summarize:}

The Hamiltonian functionals $\tilde{Q}$ that generate HK flows compatible with
rotational and gauge invariance are linear in $\Psi$ and in its adjoint
$\Psi^{\dagger}$ which implies --- using eq.(\ref{multiv deriv c}) --- that
Hamilton's equation is linear too,
\begin{equation}
\frac{d\Psi_{x}}{d\lambda}=\frac{\delta\tilde{Q}}{\delta i\hbar\Psi
_{x}^{\dagger}}~\quad\text{or}\quad i\hbar\frac{d\Psi_{x}}{d\lambda}=\int
dx^{\prime}\hat{Q}_{xx^{\prime}}\Psi_{x^{\prime}}~. \label{HK flow b}%
\end{equation}

\section{The e-Hamiltonian and the Pauli equation}

\label{H and P eq} We have shown that the condition for the simplest form of
dynamics --- one that preserves the metric, symplectic, and complex structures
--- is a Hamilton-Killing flow generated by a Hamiltonian $\tilde{H}$ of the
form (\ref{Q funct}),
\begin{equation}
\tilde{H}[\Psi,\Psi^{\dagger}]=\int\,dx_{1}dx_{2}\langle\Psi_{x_{1}}^{\dagger
}\hat{H}_{x_{1}x_{2}}\Psi_{x_{2}}\rangle_{0}~. \label{H funct a}%
\end{equation}
In ED the clock that defines entropic time is provided by the system itself
and it is natural to demand that $\tilde{H}$ --- the generator of time
translations --- be defined in terms of the very same clock. Thus, the last
ingredient in the construction of $\tilde{H}$ is to require its agreement with
(\ref{Hamiltonian a}) in order to reproduce the entropic evolution of $\rho$
given by the continuity eq.(\ref{ED c}). To proceed we introduce
\begin{equation}
\tilde{H}_{0}=\int dx\,\langle\Psi^{\dagger}\hat{H}_{0}\Psi\rangle_{0}%
\quad\text{with}\quad\,\hat{H}_{0}\Psi=\frac{1}{2m}\left(  \frac{\hbar}{i}%
\vec{\partial}-\frac{q}{c}\vec{A}\right)  ^{2}\Psi~, \label{H0 a}%
\end{equation}
and use the identity
\begin{equation}
\rho\frac{1}{2}mv^{2}=\left\langle \Psi^{\dag}\hat{H}_{0}\Psi\right\rangle
_{0}+\frac{\hbar^{2}}{2m}\rho^{1/2}\partial^{2}\rho^{1/2}-\frac{\hbar^{2}}%
{8m}\rho(\partial_{a}\,\vec{s})^{2}+\frac{\hbar q}{2mc}\rho\vec{B}\cdot\vec
{s}~, \label{identity}%
\end{equation}
derived in Appendix C. This allows us to rewrite $\tilde{H}[\rho,\Phi
,\phi,\theta]$ in (\ref{Hamiltonian a}) as
\begin{equation}
\tilde{H}[\Psi,\Psi^{\dagger}]=\int dx\,\langle\Psi^{\dagger}\hat{H}_{0}%
\Psi\rangle_{0}+F^{\prime}[\rho,\phi,\theta]~,\, \label{H funct b}%
\end{equation}
where
\begin{equation}
F^{\prime}[\rho,\phi,\theta]=F[\rho,\phi,\theta]+\int dx\,\rho\left(
\frac{\hbar^{2}}{2m}\frac{\partial^{2}\rho^{1/2}}{\rho^{1/2}}-\frac{\hbar^{2}%
}{8m}(\partial_{a}\,\vec{s})^{2}+\frac{\hbar q}{2mc}\vec{B}\cdot\vec
{s}\right)  ~. \label{F prime a}%
\end{equation}
Two conditions are to be imposed on $F^{\prime}[\rho,\phi,\theta]$: first, in
order for $\tilde{H}[\Psi,\Psi^{\ast}]$ to generate an HK flow we require that
$F^{\prime}$ itself be of the form (\ref{Q funct a}),%
\begin{equation}
F^{\prime}[\rho,\phi,\theta]=\int\,dx_{1}dx_{2}\langle\Psi_{x_{1}}^{\dagger
}\hat{V}_{x_{1}x_{2}}\Psi_{x_{2}}\rangle_{0} \label{F prime b}%
\end{equation}
for some Hermitian kernel $\hat{V}_{x_{1}x_{2}}$. And second, to reproduce the
ED flow given by (\ref{ED c}) $F^{\prime}[\rho,\phi,\theta]$ must be
independent of $\Phi$,
\begin{equation}
\frac{\delta F^{\prime}[\rho,\phi,\theta]}{\delta\Phi_{x}}=0~.
\label{F prime c}%
\end{equation}
Note that $\vec{s}$ in eq.(\ref{spin vector}) depends only on $\theta$ and
$\phi$ so the integral term in (\ref{F prime a}) is indeed independent of
$\Phi=-\hbar\bar{\chi}/2$. Substituting (\ref{spinor psi c}) into
(\ref{F prime b}) we get
\begin{equation}
F^{\prime}=\int\,dx_{1}dx_{2}\rho_{x_{1}}^{1/2}\rho_{x_{2}}^{1/2}\langle
e^{i(\Phi_{x_{2}}-\Phi_{x_{1}})/\hbar}U_{x_{1}}^{\dagger}\hat{V}_{x_{1}x_{2}%
}U_{x_{2}}(1+\vec{e}_{3})\rangle_{0} \label{F prime d}%
\end{equation}
To satisfy (\ref{F prime c}) for arbitrary choices of $\Phi_{x_{1}}$%
,$\Phi_{x_{2}}$, $U_{x_{1}}$, and $U_{x_{2}}$ we must impose that the kernel
$\hat{V}_{x_{1}x_{2}}$ be local in $x$, that is,
\begin{equation}
\hat{V}_{x_{1}x_{2}}=\delta_{x_{1}x_{2}}\hat{V}_{x_{1}}=\delta_{x_{1}x_{2}%
}(\alpha_{x_{1}}+\vec{a}_{x_{1}}+i\vec{b}_{x_{1}}+i\beta_{x_{1}})~~,
\label{V local}%
\end{equation}
where the multivector field $\hat{V}_{x_{1}}$ has been expressed in terms of
its pure grade components. Substituting into (\ref{F prime d}),
\begin{equation}
F^{\prime}=\int\,dx\rho_{x}\langle\hat{V}_{x}U_{x}(1+\vec{e}_{3}%
)U_{x}^{\dagger}\rangle_{0}=\int\,dx\rho_{x}\langle(\alpha_{x}+\vec{a}%
_{x}+i\vec{b}_{x}+i\beta_{x})(1+\vec{s}_{x})\rangle_{0}~,
\end{equation}
leads to
\begin{equation}
F^{\prime}=\int\,dx\rho_{x}(\alpha_{x}+\vec{s}_{x}\cdot\vec{a}_{x})~.
\label{F prime e}%
\end{equation}
Since the bivector $i\vec{b}$ and pseudoscalar $i\beta$ components do not
contribute they can be set to zero. Then, using (\ref{V local}) in
(\ref{F prime b}) we get
\begin{equation}
F^{\prime}=\int\,dx\langle\Psi_{x}^{\dagger}(V_{x}+\vec{a}_{x})\Psi_{x}%
\rangle_{0}~, \label{F prime g}%
\end{equation}
where we adopt the standard notation for the scalar potential, $\alpha
_{x}=V_{x}$. Substituting eqs.(\ref{H0 a}) and (\ref{F prime g}) into
(\ref{H funct b}) gives the final expression for the e-Hamiltonian,
\begin{equation}
\tilde{H}[\Psi,\Psi^{\dagger}]=\int\,dx\langle\Psi^{\dagger}\hat{H}\Psi
\rangle_{0}~ \label{H funct c}%
\end{equation}
where
\begin{equation}
\hat{H}=\hat{H}_{0}+V+\vec{a}\ .\, \label{H hat a}%
\end{equation}
Below, in section \ref{other interactions}, we discuss the effect of the
vector potential $\vec{a}$; for now we shall set $\vec{a}=0$.

We are now ready to complete the derivation of the Pauli equation. From
(\ref{HK flow b}),
\begin{equation}
\partial_{t}\Psi=\{\Psi,\tilde{H}\}\quad\text{or}\quad i\hbar\partial_{t}%
\Psi=\frac{\delta\tilde{H}}{\delta\Psi^{\dagger}}=\hat{H}\Psi~.
\label{Pauli eq a}%
\end{equation}
Then the Pauli equation is
\begin{equation}
i\hbar\partial_{t}\Psi=\frac{1}{2m}\left(  \frac{\hbar}{i}\vec{\partial}%
-\frac{q}{c}\vec{A}\right)  ^{2}\Psi+V\Psi~, \label{Pauli eq b}%
\end{equation}
which can be rewritten in its more standard form
\begin{equation}
i\hbar\partial_{t}\Psi=\frac{1}{2m}\left(  \frac{\hbar}{i}\vec{\partial}%
-\frac{q}{c}\vec{A}\right)  \cdot\left(  \frac{\hbar}{i}\vec{\partial}%
-\frac{q}{c}\vec{A}\right)  \Psi+V\Psi-\frac{\hbar q}{2mc}\vec{B}\Psi~.
\label{Pauli eq c}%
\end{equation}
The last term represents the \textquotedblleft anomalous\textquotedblright%
\ Zeeman interaction of the magnetic field with the spin. The standard
practice is to represent the basis vectors $\vec{e}_{a}$ by Pauli matrices
$\hat{\sigma}_{a}$. Then the matrix representation of the Zeeman term for the
electron ($q=-e$) is
\begin{equation}
-\frac{\hbar q}{2mc}\vec{B}\Psi\sim g\frac{e}{2mc}\frac{\hbar}{2}\hat{\sigma
}_{a}B_{a}\Psi\quad\text{with}\quad g=2\ .
\end{equation}
which exhibits the correct gyromagnetic factor.

\paragraph*{The action ---}

Once the canonical framework has been established with a Hamiltonian and
Poisson brackets it is possible to summarize the whole formalism in terms of
an action principle. The idea is to reverse the usual procedure and construct
the action so that it reproduces the desired equations. Define the
differential
\begin{equation}
\delta\mathcal{A}=\int_{t_{1}}^{t_{2}}dt\,\int_{R}dx\,\left\langle \delta
\Psi^{\dagger}\left(  i\hbar\partial_{t}\Psi-\hat{H}\Psi\right)  -\left(
i\hbar\partial_{t}\Psi^{\dagger}+\Psi^{\dagger}\hat{H}\right)  \delta
\Psi\right\rangle _{0}%
\end{equation}
with the fields $\delta\Psi^{\dagger}$ and $\delta\Psi$ vanishing at the
boundary and then integrate to get the action
\begin{equation}
\mathcal{A}=\int_{t_{1}}^{t_{2}}dt\,\int_{R}dx\,\left\langle i\hbar
\Psi^{\dagger}\partial_{t}\Psi-\Psi^{\dagger}\hat{H}\Psi\right\rangle _{0}\ .
\label{action a}%
\end{equation}
By construction, imposing $\delta\mathcal{A}=0$ leads to the Pauli equation
(\ref{Pauli eq a}). In the ED framework this construction is a convenient way
to summarize the content of the theory but it is not particularly fundamental.

\section{Hilbert space}

The formulation of the ED of a spin-1/2 particle is now complete and it is
noteworthy that the notion of Hilbert spaces did not turn out to be necessary.
However, while strictly unnecessary in principle, the introduction of Hilbert
spaces is nevertheless very convenient for calculational purposes. 

A point $\Psi\in T^{\ast}\mathbf{C}$ deserves to be called a \emph{state} in
the sense that it codifies the information required to serve as the initial
condition for a HK flow. A central feature of the HK flows (\ref{HK flow b})
or Pauli's equation (\ref{Pauli eq a}) is its linearity which implies that
given two states $\Psi_{1}$ and $\Psi_{2}$ a superposition $\Psi_{3}%
=\alpha_{1}\Psi_{1}+\alpha_{2}\Psi_{2}$ is also a state \emph{provided} the
constants $\alpha_{1}$ and $\alpha_{2}$ are such that $\Psi_{3}$ is properly
normalized. Thus normalization imposes a severe limitation on the usefulness
of linearity. The limitation can be evaded by working in the flat embedding
space $T^{\ast}\mathbf{C}^{+}$ which allows superpositions with arbitrary
constants $\alpha_{1}$ and $\alpha_{2}$ and turns the points $\Psi\in T^{\ast
}\mathbf{C}^{+}$ into vectors. This is a structure that turns out to be very useful.

\paragraph*{The inner product ---}

To take full advantage of linearity we would like to endow the space $T^{\ast
}\mathbf{C}^{+}$ with an inner product and turn it into a Hilbert space. The
idea is that tensors such as $G$ and $\Omega$ are supposed to act on
\emph{vectors} $d/d\lambda$; in principle their action on the \emph{points}
$\Psi$ is not defined. But here we deal with the special case: the space
$T^{\ast}\mathbf{C}^{+}$ is flat and its tangent spaces are identical to the
space itself. Therefore, vectors tangent to the manifold and points on the
manifold are objects of the same kind. In other words, since $T^{\ast
}\mathbf{C}^{+}$ is flat the tensors $G$ and $\Omega$ can act on points $\Psi$
too. The choice of an inner product for the points $\Psi$ is, then, natural in
the sense that the necessary ingredients --- the symplectic form $\Omega$ and
the metric tensor $G$ --- are already available. 

We can at this point introduce the Dirac notation to represent the spinor wave
functions $\Psi$, eq.(\ref{spinor psi d}), as vectors $|\Psi\rangle$ in a
Hilbert space. The scalar product $\langle\Psi_{1}|\Psi_{2}\rangle$ is defined
by
\begin{equation}
\langle\Psi_{1}|\Psi_{2}\rangle\overset{\text{def}}{=}\left(  G_{\mu x,\nu
x^{\prime}}+\frac{i}{2\hbar}\Omega_{\mu x,\nu x^{\prime}}\right)  \Psi^{\mu
x}\Psi^{\nu x^{\prime}}~
\end{equation}
where
\begin{equation}
G_{\mu x,\nu x^{\prime}}+\frac{i}{2\hbar}\Omega_{\mu x,\nu x^{\prime}}%
=\frac{1}{i\hbar}%
\begin{bmatrix}
0 & 0 & 0 & 0\\
1 & 0 & 0 & 0\\
0 & 0 & 0 & 0\\
0 & 0 & 1 & 0
\end{bmatrix}
~.
\end{equation}
A quick calculation leads to the familiar result
\begin{equation}
\langle\Psi_{1}|\Psi_{2}\rangle=\int dx\left(  \psi_{1+x}^{\ast}\psi
_{2+x}+\psi_{1-x}^{\ast}\psi_{2-x}\right)  ~.
\end{equation}
The map $\Psi\longleftrightarrow$ $|\Psi\rangle$ is defined by
\begin{equation}
|\Psi\rangle=\int dx\,\left(  \psi_{+}|u_{+}\rangle|x\rangle+\psi_{-}%
|u_{-}\rangle|x\rangle\right)  ~,
\end{equation}
where the spinor basis $\{|u_{+}\rangle,|u_{-}\rangle\}$ and the position
basis\ $\{|x\rangle\}$ are both orthogonal and complete.

Note also that the mixed tensor $J$, which is normally meant to act on tangent
vectors, is now also free to act on the points $\Psi$. The action of $J$,
eq.(\ref{J tensor}), on a spinor wave function $\Psi$, eq.(\ref{spinor psi e}%
), is
\begin{equation}
J^{\mu x}{}_{\nu x^{\prime}}\Psi^{\nu x^{\prime}}=%
\begin{bmatrix}
i & 0 & 0 & 0\\
0 & -i & 0 & 0\\
0 & 0 & i & 0\\
0 & 0 & 0 & -i
\end{bmatrix}
\,%
\begin{bmatrix}
\psi_{+x}\\
i\hbar\psi_{+x}^{\ast}\\
\psi_{-x}\\
i\hbar\psi_{-x}^{\ast}%
\end{bmatrix}
=%
\begin{bmatrix}
i\psi_{+x}\\
i\hbar(i\psi_{+x})^{\ast}\\
i\psi_{-x}\\
i\hbar(i\psi_{-x})^{\ast}%
\end{bmatrix}
~.
\end{equation}
which shows that $J$ plays the role of multiplication by $i$, that is,
$\psi_{\pm}\overset{J}{\rightarrow}i\psi_{\pm}$.

The bilinear Hamilton functionals $\tilde{Q}[\Psi,\Psi^{\ast}]$ with kernel
$\hat{Q}(x,x^{\prime})$ in eqs.(\ref{Q funct a}) or (\ref{Q funct b}) can now
be written in terms of a Hermitian operator $\hat{Q}$ and its matrix
elements,
\begin{equation}
\tilde{Q}[\Psi,\Psi^{\ast}]=\langle\Psi|\hat{Q}|\Psi\rangle\quad
\text{and}\quad\hat{Q}_{\sigma\sigma^{\prime}}(x,x^{\prime})=\langle
u_{\sigma}|\langle x|\hat{Q}|u_{\sigma^{\prime}}\rangle|x^{\prime}\rangle~.
\end{equation}
The corresponding Hamilton-Killing flows are given by
\begin{equation}
i\hbar\frac{d}{d\lambda}|\Psi\rangle=\hat{Q}|\Psi\rangle~.
\end{equation}
These flows are described by unitary transformations
\begin{equation}
|\Psi(\lambda)\rangle=\hat{U}_{Q}(\lambda)|\Psi(0)\rangle\quad\text{where}%
\quad\hat{U}_{Q}(\lambda)=\exp\left(  -\frac{i}{\hbar}\hat{Q}\lambda\right)
~.
\end{equation}

\section{Discussion and interpretation}

The discussion so far has been directed towards reconstructing and therefore
justifying the mathematical formalism behind the Pauli equation but there is a
second goal of clarifying the interpretation of QM and, in particular, of
spin. Clearly the formalism we have derived is empirically equivalent to the
usual Pauli equation. Next we turn our attention to the second task which, as
we shall see, yields a physical picture that departs significantly from the
standard Copenhagen, Bohmian, and the Hestenes interpretations.

\subsection{Energy}

The e-Hamiltonian $\tilde{H}$ is the generator of translations in time. It
tells us how wave functions evolve and, therefore, it tells us that
\emph{energy is an epistemic concept} associated to the wave function and not
to the particle. To gain further insight substitute (\ref{Pauli eq a}) into
(\ref{H funct c}),
\begin{equation}
\tilde{H}=\int\,dx\langle\Psi^{\dagger}i\hbar\partial_{t}\Psi\rangle
_{0}~.~\label{H funct d}%
\end{equation}
Using (\ref{spinor psi b}) a calculation similar to (\ref{w dot s d}) gives%
\begin{equation}
\langle\Psi^{\dagger}i\hbar\partial_{t}\Psi\rangle_{0}=\rho\,\vec{\omega}%
_{t}\cdot\vec{S}\label{energy dens a}%
\end{equation}
where a factor $\hbar/2$ has been absorbed into the spin vector to obtain the
more familiar normalization%
\begin{equation}
\vec{S}(x)=\frac{\hbar}{2}\vec{s}(x)\quad\text{and}\quad\vec{\omega}_{t}%
=\vec{e}_{3}\partial_{t}\phi+\vec{e}_{\phi}\partial_{t}\theta+\vec
{s}\,\partial_{t}\bar{\chi}\label{spin vector b}%
\end{equation}
is the angular velocity of the spin frame. Eq.(\ref{energy dens a}) allows us
to define a \emph{local energy},
\begin{equation}
\varepsilon(x)\overset{\text{def}}{=}\,\vec{\omega}_{t}(x)\cdot\vec
{S}(x)=-\partial_{t}\Phi+\frac{\hbar}{2}\cos\theta\,\partial_{t}\phi~.
\end{equation}
Thus, the local energy is basically just the rate of rotation of the spin
frame about the spin vector. Furthermore, for those special states $\Psi_{E}$
that happen to be eigenstates of $\hat{H}$,
\begin{equation}
\hat{H}\Psi_{E}=E\Psi_{E}~,
\end{equation}
eq.(\ref{H funct c}) gives
\begin{equation}
\tilde{H}[\Psi_{E},\Psi_{E}^{\dagger}]=\int\,dx\,\rho E=E\quad\text{so that
}\varepsilon(x)=E~.
\end{equation}
Thus, energy eigenstates are those for which the local energy $\varepsilon(x)$
is independent of position.

Our concept of a local energy bears an inevitable formal resemblance to the
corresponding \textquotedblleft local observable\textquotedblright\ proposed
by Hestenes and Gurtler \cite{Hestenes Gurtler 1971}\ but the conceptual
difference is significant: our local energy is an epistemic object, a property
of the wave function; their local energy is ontic, a property of the particle.

\subsection{Linear momentum}

Just as with energy, the formalism of momentum as the generator of spatial
translations is very familiar, but the physical picture is new and perhaps
surprising. Under a translation $x^{a}\rightarrow x_{\xi}^{a}=x^{a}+\xi^{a}$
the wave function transforms as
\begin{equation}
\Psi(x)\rightarrow\Psi_{\xi}(x_{\xi})=\Psi(x)\quad\text{or}\quad\Psi_{\xi
}(x)=\Psi(x_{-\xi})
\end{equation}
and the momentum functional $\tilde{P}_{a}$ is defined so that
\begin{equation}
\delta_{\xi}\Psi=\Psi_{\xi}(x)-\Psi(x)=-\xi^{a}\partial_{a}\Psi=\{\Psi
,\tilde{P}_{a}\xi^{a}\}~.
\end{equation}
The required $\tilde{P}_{a}$ is such that
\begin{equation}
\frac{\delta\tilde{P}_{a}}{\delta i\hbar\Psi^{\dagger}}=-\partial_{a}\Psi~,
\end{equation}
which, with the help of (\ref{multiv deriv c}), can be integrated to
\begin{equation}
\tilde{P}_{a}[\Psi,\Psi^{\dagger}]=\int\,dx\,\langle\Psi^{\dagger}\frac{\hbar
}{i}\partial_{a}\Psi\rangle_{0}~.\label{P funct a}%
\end{equation}
Further insight comes from a calculation similar to (\ref{w dot s d}) which
relates the momentum density to the spin,%
\begin{equation}
\langle\Psi^{\dagger}\frac{\hbar}{i}\partial_{a}\Psi\rangle_{0}=-\rho
\,\vec{\omega}_{a}\cdot\vec{S}~,\label{mom density}%
\end{equation}
where
\begin{equation}
\vec{\omega}_{a}=\vec{e}_{3}\partial_{a}\phi+\vec{e}_{\phi}\partial_{a}%
\theta+\vec{s}\,\partial_{a}\bar{\chi}~.\label{wa}%
\end{equation}
Eq.(\ref{mom density}) allows us to define a \emph{local} \emph{momentum},
\begin{equation}
p_{a}(x)\overset{\text{def}}{=}\,-\vec{\omega}_{a}(x)\cdot\vec{S}%
(x)~,\label{local momentum a}%
\end{equation}
which, using eq.(\ref{drift c}), yields the familiar expression%
\begin{equation}
p_{a}(x)=mv_{a}(x)+\frac{q}{c}A_{a}(x)~.\label{local momentum b}%
\end{equation}
Thus the local momentum is an epistemic quantity --- a property of the wave
function --- that can be inferred but is not itself directly observable.
Further insight is obtained using (\ref{wa}) and (\ref{local momentum a}) to
write
\begin{equation}
p_{a}=\partial_{a}\Phi-\frac{\hbar}{2}\cos\theta\,\partial_{a}\phi
~,\label{local momentum c}%
\end{equation}
which shows that the local momentum depends only on spatial derivatives of the
phase $\Phi$ and the spin and not on the probability density. This might at
first look surprising. It implies, for example, that in the ground state of
hydrogen, where both $\Phi$ and $\vec{S}$ are uniform in space, the local
momentum $p_{a}$ and the velocity $v^{a}$ vanish. \emph{There is a zero point
energy but no zero point motion}. Is this a violation of the uncertainty
principle? No. Just as for energy, the momentum eigenstates are those for
which the local momentum (\ref{local momentum b}) is independent of position.
It turns out that the outcomes of what are normally called \textquotedblleft
measurements of momentum\textquotedblright\ do not reflect the pre-existing
values of local momentum; they reflect eigenvalues that are actually
\textquotedblleft created\textquotedblright\ during the measurement process
and are distributed according to the familiar Born rule from the Fourier
transform of the wave function \cite{Johnson Caticha 2011}\cite{Vanslette
Caticha 2016}. And how is it that an electron at rest does not immediately
start falling towards the nucleus? What holds it in place? The answer is that
the electron does not obey Newtonian dynamics and our intuition must be
re-educated. In the ED approach there are no forces acting on the electron. ED
is a dynamics of probabilities; what is being pushed is not the particle but
our expectations about where the electron will be found.

\subsection{Orbital and spin angular momentum}

We finally come to the interpretation of spin. First the familiar part:
angular momentum is the generator of rotations. Under a rotation by an angle
$\xi$ about the axis $\vec{n}$ we have
\begin{equation}
\vec{x}\rightarrow\vec{x}_{\xi}=R_{\xi}\vec{x}R_{\xi}^{\dagger}\quad
\text{where}\quad R_{\xi}=e^{-i\vec{n}\xi/2}%
\end{equation}
and the wave function transforms as
\begin{equation}
\Psi(x)\rightarrow\Psi_{\xi}(x_{\xi})=R_{\xi}\Psi(x)\quad\text{or}\quad
\Psi_{\xi}(x)=R_{\xi}\Psi(x_{-\xi})~.
\end{equation}
In a rotation by the infinitesimal angle $\xi$,
\begin{equation}
\vec{x}_{-\xi}=\vec{x}-\xi\,\vec{n}\times\vec{x}~,
\end{equation}
and the change of the wave function is%
\begin{equation}
\delta_{\xi}\Psi(x)=R_{\xi}\Psi(x_{-\xi})-\Psi(x)=-\frac{i}{\hbar}\xi
\,n^{a}\hat{J}_{a}\Psi(x)
\end{equation}
where the total angular momentum operator,
\begin{equation}
\hat{J}_{a}=\hat{L}_{a}+\hat{S}_{a}~,\label{ang mom a}%
\end{equation}
includes orbital and spin contributions,%
\begin{equation}
\hat{L}_{a}=\vec{e}_{a}\cdot(\vec{x}\times\frac{\hbar}{i}\vec{\partial}%
)\quad\text{and}\quad\hat{S}_{a}=\frac{\hbar}{2}\vec{e}_{a}~.\label{ang mom b}%
\end{equation}

The angular momentum functional $\tilde{J}_{a}$ generates rotations according
to the Poisson bracket
\begin{equation}
\delta_{\xi}\Psi=\{\Psi,\xi\,n^{a}\tilde{J}_{a}\}\quad\text{so that}\quad
\frac{\delta\tilde{J}_{a}}{\delta i\hbar\Psi^{\dagger}}=\hat{J}_{a}\Psi\,,~
\end{equation}
which, using (\ref{multiv deriv c}) can be integrated to%
\begin{equation}
\tilde{J}_{a}[\Psi,\Psi^{\dagger}]=\int\,dx\,\langle\Psi^{\dagger}\hat{J}%
_{a}\Psi\rangle_{0}=\int\,dx\,\langle\Psi^{\dagger}\left(  \frac{\hbar}%
{i}\varepsilon_{abc}x_{b}\partial_{c}+\frac{\hbar}{2}\vec{e}_{a}\right)
\Psi\rangle_{0}~.\label{ang mom c}%
\end{equation}

To gain further insight into spin we note that in a Pauli matrix
representation of GA, $\vec{e}_{a}\sim\hat{\sigma}_{a}$, one recovers the
familiar expression for the spin operator, $\hat{S}_{a}\sim(\hbar
/2)\hat{\sigma}_{a}$. Furthermore, a calculation similar to (\ref{w dot s d})
allows us to define the local spin density,%
\begin{align}
\langle\Psi^{\dagger}\frac{\hbar}{2}\vec{e}_{a}\Psi\rangle_{0} &  =\frac
{\hbar}{2}\rho\,\langle\vec{e}_{a}Uu_{+}u_{+}^{\dagger}U^{\dagger}\rangle
_{0}=\frac{\hbar}{2}\rho\,\langle\vec{e}_{a}(1+\vec{s})\rangle_{0}\nonumber\\
&  =\frac{\hbar}{2}\rho\,\vec{e}_{a}\cdot\vec{s}=\rho\,S_{a}%
~,\label{local spin density}%
\end{align}
which shows that the vector spin functional is
\begin{equation}
\tilde{S}[\Psi,\Psi^{\dagger}]=\vec{e}_{a}\int\,dx\,\langle\Psi^{\dagger}%
\frac{\hbar}{2}\vec{e}_{a}\Psi\rangle_{0}=\int\,dx\,\rho\,\vec{S}%
~,\label{Stilde}%
\end{equation}
where $\vec{S}(x)$ is the local spin vector. 

\paragraph*{What is spinning? ---}

Once an angular momentum such as spin is identified it is tempting to ask
\textquotedblleft what is it that rotates?\textquotedblright\ But this is a
classical prejudice that must be avoided. Angular momentum is a mathematical
abstraction; it is the generator of rotations but the rotations need not be
dynamical. In particular, there need not exist any small rigid rotators or
vortices in some underlying fluid. The answer is that \emph{nothing is
spinning.} Once again, the ground state of hydrogen, with $\vec{v}=0$ and a
constant $\vec{S}$, is instructive. This is a situation in which neither the
electron is moving nor are the probabilities flowing and yet the integral in
(\ref{Stilde}) gives a total spin of $\tilde{S}=\hbar/2$.

\paragraph*{The magnitude of spin? ---}

The spin vector $\vec{S}$ has magnitude $\hbar/2$. Note however that the GA
expression
\begin{equation}
\hat{S}_{1}^{2}+\hat{S}_{2}^{2}+\hat{S}_{3}^{2}=(\frac{\hbar}{2})^{2}\left(
\vec{e}_{1}^{2}+\vec{e}_{2}^{2}+\vec{e}_{3}^{2}\right)  =\frac{3}{4}\hbar^{2}%
\end{equation}
corresponds to
\begin{equation}
\hat{S}^{2}\sim(\frac{\hbar}{2})^{2}\left(  \hat{\sigma}_{1}^{2}+\hat{\sigma
}_{2}^{2}+\hat{\sigma}_{3}^{2}\right)
\end{equation}
and exactly reproduces the well-known expression
\begin{equation}
\hat{S}^{2}\sim\hbar^{2}s(s+1)\mathbf{\hat{1}}\quad\text{with}\quad s=\frac
{1}{2}~.~
\end{equation}
But the interpretations are completely different: the latter is meant to
represent the magnitude squared of the spin operator while the former is just
the square of the diagonal of a cube of side $\hbar/2$.

\subsection{The electric current and the magnetic dipole moment}

When quantities such as mass or charge are interpreted as ontic properties of
the particle it is only natural to expect that the corresponding densities and
currents be closely related. For example, if the particle flux is $\rho\vec
{v}$ one expects the fluxes of mass and charge to be $m\rho\vec{v}$ and
$q\rho\vec{v}$ --- each moving particle automatically carries an amount $m$ of
mass and $q$ of charge. Conversely, when mass and charge are not ontic but
epistemic properties assigned to the wave function the question of whether the
mass and charge currents coincide must be revisited.

We define the electric charge and electric current through their coupling to
the electromagnetic potentials. This is most easily implemented in terms of
the action, eq.(\ref{action a}),
\begin{equation}
\mathcal{A}=\int dt\,dx\,\left\langle \Psi^{\dagger}\left[  i\hbar\partial
_{t}-\frac{1}{2m}\left(  \frac{\hbar}{i}\vec{\partial}-\frac{q}{c}\vec
{A}\right)  ^{2}-qA^{0}\right]  \Psi\right\rangle \label{action b}%
\end{equation}
where the potential $V$ is written as $qA^{0}$. The electric charge $\rho_{e}$
and current density $\vec{J}_{e}$ are defined as functional derivatives with
respect to the potentials $A^{0}$ and $A^{a}$,
\begin{equation}
\delta\mathcal{A}\overset{\text{def}}{=}\int dt\,dx\,\left[  -\rho_{e}\delta
A^{0}+\frac{1}{c}\vec{J}_{e}\cdot\delta\vec{A}\right]
\end{equation}
A straightforward calculation gives the charge density
\begin{equation}
\rho_{e}=q\langle\Psi^{\dagger}\Psi\rangle=q\rho
\end{equation}
and an electric current that splits into convective and spin components,%

\begin{equation}
J_{e}^{a}=\frac{q}{m}\langle\Psi^{\dagger}(\frac{\hbar}{i}\partial^{a}%
-\frac{q}{c}A^{a})\Psi\rangle+\frac{q}{m}\varepsilon^{abc}\partial_{b}%
\langle\Psi^{\dagger}\frac{\hbar}{2}\vec{e}_{c}\Psi\rangle~.
\end{equation}
Using (\ref{pr current}) and (\ref{local spin density}) this is written as
\begin{equation}
\vec{J}_{e}=q\rho\vec{v}+c\,\vec{\partial}\times\vec{M}~,
\end{equation}
which explicitly shows that the spin contributes to the electric current as a
magnetization current with a magnetization $\vec{M}$ proportional to the spin
density $\rho\vec{S}$,
\begin{equation}
\vec{M}=\frac{q}{mc}\rho\vec{S}~.
\end{equation}

The convective current $q\rho\vec{v}$ and the magnetization current
$c\,\vec{\partial}\times\vec{M}$ are quite independent. In the ground state of
hydrogen, with $\vec{v}=0$ and $\vec{S}$ constant, nothing is moving but it is
easy to check that the electric current does not vanish. If we take $\vec{S}$
in the $\vec{e}_{3}$ direction then the current forms a toroidal ring in the
vicinity of the $xy$ plane. It is this \textquotedblleft
current\textquotedblright\ in which nothing flows that is responsible for the
electron's magnetic dipole moment. Alternatively, we can adopt a manifestly
\textquotedblleft static\textquotedblright\ model and explain the dipole
moment as arising from a magnetic pole density $-\vec{\partial}\cdot\vec{M}$.
Again, it is easy to check that this gives the correct North and South pole
densities concentrated around the lower and upper $\vec{e}_{3}$ axis respectively.

\subsection{What happens in a spin measurement?}

Since position is the only ontic quantity all measurements in ED must be
traced back to detecting the presence or absence of a particle within the
volume of a device. Position is the only beable and therefore the only
observable. All other quantities, such as momentum or energy, are necessarily
epistemic and cannot therefore be observed. These quantities can at best be
inferred from position detections and, accordingly, it is appropriate to refer
to them as \emph{inferables}.\footnote{The term `inferable' is due to K.
Vanslette. With the recent development of techniques to perform weak
measurements the wave function is itself is an inferable. (See \emph{e.g.},
\cite{Vanslette Caticha 2016} and references therein.)} Spin is no exception.

The process of \textquotedblleft measuring\textquotedblright\ spin by means of
a Stern-Gerlach (SG) experiment is well known. The central question of what
precisely happens during the experiment depends on the particular
interpretation of quantum mechanics one adopts. (For a recent review with
references to the literature see \cite{Piceno et al 2018}.)

ED resembles Bohmian mechanics in that particles have definite positions and
its formalism includes functions $(\Phi,\theta,\phi)$ that play the role of a
pilot wave. However, the differences are considerable. Bohmian mechanics
attempts to provide a complete description of reality. The wave function
$\Psi$ is a real field that lives in configuration space and acts on the
particles without the particles reacting back upon it --- peculiarities that
have stood in the way of a wider acceptance of the Bohmian interpretation. In
contrast, ED's pragmatic goal is limited to make the best possible inferences
on the basis of very incomplete information; the ED variables $(\Phi
,\theta,\phi)$ are purely epistemic and there is no implication that the
particles are carried by a pilot wave or pushed by any other force. In fact ED
is silent on the issue of what if anything makes the particles move as they
do. The wave function $\Psi$ exerts no causal influence on the particles
themselves; what the wave function $\Psi$ does is to guide our expectations of
where and when the particles will be. Having described the differences between
the ED and Bohmian frameworks we now proceed to exploit the similarities and
borrow from the analysis of the Pauli equation by Dewdney \emph{et al.}
\cite{Dewdney et al 1986}.

The particle is prepared in a state that describes an initial wave packet
$\Psi(t_{0},\vec{x})$ with a spin vector $\vec{S}(t_{0})=\hbar\vec{s}/2$ in
some definite but arbitrary direction $\vec{s}$. Solving the Pauli equation
shows that the particle can follow any of a congruence of smooth trajectories.
Which trajectory is actually followed depends only on the particle's initial
position within the wave packet $\Psi(t_{0},\vec{x})$. As the particle
traverses the SG\ device the spin vector field $\vec{S}(t,\vec{x})$ evolves
continuously. The analysis of Dewdney \emph{et al.} shows that the wave packet
splits into two separate components along the direction $\vec{e}_{z}$ of the
inhomogeneous magnetic field. Whether the particle ends in the upper or the
lower wave packet depends only which trajectory was followed.

The remarkable result is that the evolution leads to a strong correlation
between final values of the spin vectors and the particle trajectories: all
those trajectories that end up in the upper wave packet will have final spin
$\vec{S}=+\hbar\vec{e}_{z}/2$ while all trajectories that end up in the lower
wave packet will have final spin $\vec{S}=-\hbar\vec{e}_{z}/2$. The analysis
also shows that the actual probabilities of being in the upper of lower wave
packet turn out to be given by the standard Born rule. (Note that these are
the probabilities of \emph{being} and not merely of being \emph{found} in one
or the other packets.)

It is only at this final stage that an actual measurement happens: the
particle's position is detected. Depending on how precise the position
measurement is this allows us not only to infer the trajectory that was
actually followed but also the final value of the spin vector being either
$+\hbar/2$ or $-\hbar/2$ along the $\vec{e}_{z}$ direction.

This result is remarkable in several ways. The discrete outcomes $\pm\hbar/2$
explain the so-called \textquotedblleft space quantization\textquotedblright%
\ without invoking the eigenvalues of $\hat{\sigma}_{z}$. Standard QM is
forced to postulate the Born rule and the rule that the possible outcomes of a
measurement are given by the eigenvalues of the Hermitian operator being
measured. Within ED such postulates are no longer needed.\footnote{Unlike ED,
in Bohmian mechanics the identification of the probability $\rho$ with
$|\Psi|^{2}$ requires for its justification additional dynamical arguments
that are highly nontrivial.} (See also \cite{Johnson Caticha 2011}%
\cite{Vanslette Caticha 2016}.) One can also see that the outcomes of this
\textquotedblleft measurement\textquotedblright\ of spin do not reflect the
initial pre-existing spin value. The outcome of the experiment was in effect
\textquotedblleft created\textquotedblright\ by the process of measurement.
Indeed, as emphasized long ago by Bell \cite{Bell 1990}, the term
`measurement' provides a very inadequate description of what actually happens
in the SG experiment.

\subsection{Other interactions}

\label{other interactions}Here we briefly comment on additional interactions
that might be included in (\ref{H hat a}) through the scalar potential $V$ or
the vector potential $\vec{a}$. The requirement that the kernel $\hat
{V}_{x_{1}x_{2}}$ be local, eq.(\ref{V local}), is very restrictive because it
rules out interactions that involve spatial derivatives of $\Psi$. Therefore
momentum- and velocity-dependent interactions --- see e.g.
eq.(\ref{local momentum c}) --- are ruled out. These include relativistic
corrections to the kinetic energy that can be described as a scalar
perturbation ($\varpropto p^{4}$) and the spin-orbit interaction. Indeed, a
magnetic dipole $\vec{\mu}\varpropto\vec{S}$ in motion exhibits an electric
dipole $\vec{d}$ which interacts with the electric field in the atom. However,
the interaction depends on the velocity $\vec{v}$,
\begin{equation}
\hat{H}_{so}\varpropto-\vec{d}\cdot\vec{E}=-(\frac{\vec{v}}{c}\times\vec{\mu
})\cdot\vec{E}~,
\end{equation}
and is therefore ruled out because it involves derivatives. In order to
reproduce these relativistic effects one would need a fully relativistic
treatment that from the start takes the positron degrees of freedom into account.

The vector potential $\vec{a}$ in (\ref{H hat a}) does, however, allow new
interactions. In the presence of an external electromagnetic field the only
available vectors are the magnetic and the electric fields, $\vec{B}$ and
$\vec{E}$, and this allows a Hamiltonian of the form
\begin{equation}
\hat{H}=\hat{H}_{0}+V+\kappa_{m}\,\vec{B}+\kappa_{e}\vec{E}\ .\,
\label{H hat b}%
\end{equation}
for some constants $\kappa_{m}$ and $\kappa_{e}$. In the usual matrix
representation $\vec{e}_{a}\sim\hat{\sigma}_{a}$, the fields $\vec{B}$ and
$\vec{E}$ are written as $B_{a}\hat{\sigma}_{a}$ and $E_{a}\hat{\sigma}_{a}$
where $\hbar\hat{\sigma}_{a}/2$ are called the spin operators. Then the term
$\kappa_{m}\,\vec{B}\varpropto\vec{S}\cdot\vec{B}$ describes the anomalous
magnetic moment of say, the proton or the neutron. The analogous electric
dipole term $\kappa_{e}\vec{E}\varpropto\vec{S}\cdot\vec{E}$ is in principle
allowed but, as we show next, its presence would signal a violation of time
reversal invariance.

\subsubsection*{Time reversal}

The transformation of the spinor wave function $\Psi$ involves taking
$t\rightarrow-t$ and the spatial inverse which plays the role of complex
conjugation. In addition, since spatial inversion has the effect of changing
the ideal,%
\begin{equation}
(1+\vec{e}_{3})^{\ast}=1-\vec{e}_{3}~,
\end{equation}
it is necessary to right multiply by $\vec{e}_{1}$ in order to restore the
original ideal,
\begin{equation}
u_{+}=\frac{1+\vec{e}_{3}}{\sqrt{2}}\rightarrow u_{+}^{T}=u_{+}^{\ast}\vec
{e}_{1}=\frac{1-\vec{e}_{3}}{\sqrt{2}}\vec{e}_{1}=\vec{e}_{1}\frac{1+\vec
{e}_{3}}{\sqrt{2}}~.
\end{equation}
(Right multiplication by $\vec{e}_{2}$ also works.) Therefore the time
reversal of $\Psi$ is implemented by%

\begin{equation}
\Psi_{t}^{T}(x)=\Psi_{-t}^{\ast}(x)\vec{e}_{1}~,
\end{equation}
or, equivalently
\begin{equation}
\Psi_{t}(x)=\psi_{+}(x,t)u_{+}+\psi_{-}(x,t)u_{-}\longrightarrow\Psi_{t}%
^{T}(x)=-\psi_{-}^{\ast}(x,-t)u_{+}+\psi_{+}^{\ast}(x,t)u_{-}~.
\end{equation}

To find the time-reversed spin vector we appeal to the \textquotedblleft
Born\textquotedblright\ rule, eq.(\ref{Born rule}) and use
\begin{equation}
\Psi^{T}\Psi^{T\dagger}=\rho U^{\ast}u_{+}^{\ast}\vec{e}_{1}\vec{e}_{1}%
u_{+}^{\ast}U^{\dagger\ast}=\rho U(1-\vec{e}_{3})U^{\dagger}=\rho(1-\vec{s})
\end{equation}
so that
\begin{equation}
\rho^{T}=\langle\Psi^{T}\Psi^{T\dagger}\rangle_{0}=\rho\quad\text{and}%
\quad\rho\vec{s}^{T}=\langle\Psi^{T}\Psi^{T\dagger}\rangle_{1}=-\rho\vec{s}~.
\end{equation}
Therefore the time-reversed spin vector is
\begin{equation}
\vec{S}_{t}^{T}(x)=-\vec{S}_{-t}(x)~.
\end{equation}

The time reversal of the Pauli equation (\ref{Pauli eq b}) involves the
time-reversed electromagnetic fields,
\begin{equation}
\vec{A}_{t}^{T}(x)=-\vec{A}_{-t}(x),\quad\vec{E}_{t}^{T}(x)=\vec{E}%
_{-t}(x),\quad\text{and}\quad\vec{B}_{t}^{T}(x)=-\vec{B}_{-t}(x)~.
\end{equation}
It is easy to check that the transformed $\Psi_{t}^{T}(x)$ is a solution of
the transformed Pauli equation (\ref{Pauli eq b}), that is, time reversal is a
symmetry. However, under $T$ an additional electric moment interaction
$\kappa_{e}\vec{E}$ as in (\ref{H hat b}) changes sign,
\begin{equation}
\kappa_{e}\vec{E}_{t}\Psi_{t}\overset{T}{\longrightarrow}(\kappa_{e}\vec
{E}_{-t}\Psi_{-t})^{\ast}\,\vec{e}_{1}=-\kappa_{e}\vec{E}_{t}^{T}\Psi_{t}%
^{T}~,
\end{equation}
and would spoil the symmetry. Thus, time reversal symmetry allows a magnetic
dipole but rules out an electric dipole.

\section{Some final comments}

We conclude with a summary of the main results. In this paper the ED framework
has been extended to describe a spin-1/2 particle. In this model the
particle's position is real while the wave function is epistemic. Position is
the only ontic variable; all other dynamical variables including spin are
properties of the wave function and therefore they are epistemic too.

The language of geometric algebra has been used to describe the rotational
properties of spin and to construct the spinor wave function as an element of
a minimal left ideal. The input of physical information is carried out by
constraints that must be continuously updated. The geometric criterion for
these updates requires identifying suitable symplectic and information
geometric structures of phase space. The ED that preserves both the symplectic
structure (a Hamiltonian flow) and the metric structure (a Killing flow) is
described by a linear Hamiltonian flow. The additional condition that the
Hamiltonian be the generator of translations in \emph{entropic} time leads to
the Pauli equation. The introduction of Hilbert spaces as an additional
structure is optional but it is very useful in that it allows us to exploit
the calculational advantages of the linearity of the Pauli equation.

Spin in ED is not a feature of a rotating body, and it is not a feature of the
motion of a point particle; it is an epistemic property of the wave function.
The spin does not guide the motion of the particle; instead it guides the
assignment of probabilities of the particle's position. It is noteworthy that
in the ED approach to QM the only probabilities are probabilities of position
and, therefore, it makes no sense to talk about an isolated spin degree of
freedom divorced from its associated particle. How would such an isolated spin
ever be measured?

We have also seen that in ED the spin $\vec{S}$ is not quantized; what turns
out to be quantized are the outcomes of Stern-Gerlach experiments. More
generally, the statements that probability, mass, momentum, charge, and
angular momentum including spin are not ontic but epistemic quantities are not
innocent. They force an extreme revision of our intuitions about physics.
Probabilities are not substances; they may change but they neither move nor
flow. And similar considerations apply to spin. Accepting that spin is an
epistemic concept forces us to abandon the classical intuition that an angular
momentum reflects the existence of some ontic substance that is actually rotating.

\subparagraph*{Acknowledgments}

We would like to thank M. Abedi, D. Bartolomeo, C. Cafaro, N. Caticha, F.X.
Costa, S. DiFranzo, K. Earle, A. Giffin, S. Ipek, D.T. Johnson, K. Knuth, O.
Lunin, S. Nawaz, P. Pessoa, M. Reginatto, C. Rodr\'{\i}guez, and K. Vanslette,
for valuable discussions on entropic inference and entropic dynamics and for
their many insights and contributions at various stages of this program.

\section*{Appendix A: Brief review of geometric algebra}

There is a vast literature devoted to the study of GA; an excellent
introduction is the classic text by David Hestenes \cite{Hestenes 1966} (see
also \cite{Doran Lasenby 2003}).

\subsection*{A.1 Multivectors}

The geometric algebra of Euclidean three-dimensional space --- the Pauli
algebra $\mathcal{G}_{3}$ --- is a graded algebra. The central concept is the
\emph{geometric product}. The geometric algebra of Euclidean three-dimensional
space --- the Pauli algebra $\mathcal{G}_{3}$ --- is a graded algebra. The
central concept is the \emph{geometric product}. The product of two vectors
$\vec{a}$ and $\vec{b}$ is defined by
\begin{equation}
\vec{a}\,\vec{b}=\vec{a}\cdot\vec{b}+\vec{a}\wedge\vec{b}~. \tag{A1}%
\label{geo prod a}%
\end{equation}
The \emph{inner}\textbf{\ }product, $\vec{a}\cdot\vec{b}=\vec{b}\cdot\vec
{a}=\delta_{ab}a^{a}b^{b}$, is the familiar scalar \emph{dot}\textbf{\ }%
product. The \emph{outer}\textbf{\ }product or \emph{wedge} product, $\vec
{a}\wedge\vec{b}=-\vec{b}\wedge\vec{a}$, is a bivector and represents the
directed area defined by $\vec{a}$ and $\vec{b}$. A generic multivector $A$
may include components of grades $0$ through $3$, and can be written as%
\begin{equation}
A=\alpha+\vec{a}+i\vec{b}+i\beta~, \tag{A2}\label{multivector}%
\end{equation}
where $\alpha$ and $\beta$ are (real) scalars and $\vec{a}$ and $\vec{b}$ are
(real) vectors. The elements of maximum grade, which in three dimensions is
grade three, are called pseudoscalars. The unit pseudoscalar is denoted $i$
because $i^{2}=-1$ and commutes with all elements in the algebra. We adopt the
standard notation of denoting the grade $k$ part by $\langle\ldots\rangle_{k}$
so that, for example, in eq.(\ref{multivector}) we have $\langle A\rangle
_{2}=i\vec{b}$.

Introducing an orthonormal basis $\{\vec{e}_{a}\}$,%
\begin{equation}
\vec{e}_{a}\cdot\vec{e}_{b}=\frac{1}{2}\left(  \vec{e}_{a}\vec{e}_{b}+\vec
{e}_{b}\vec{e}_{a}\right)  =\delta_{ab}~, \tag{A3}\label{Pauli a}%
\end{equation}
allows a generic vector $\vec{a}$ to be expressed as $\vec{a}=a^{a}\vec{e}%
_{a}$ and the pseudoscalar is
\begin{equation}
\vec{e}_{1}\vec{e}_{2}\vec{e}_{3}=i\quad\text{or}\quad\vec{e}_{a}\vec{e}%
_{b}\vec{e}_{c}=i\varepsilon_{abc}~. \tag{A4}\label{pseudoscalar}%
\end{equation}
In three dimensions a generic bivector $B$ can always be written in the form
of an \textquotedblleft imaginary\textquotedblright\ vector, $B=i\vec{b}$,
which serves to define the familiar vector product, $\vec{a}\times\vec{b}$,
as
\begin{equation}
\vec{a}\wedge\vec{b}=i\,\vec{a}\times\vec{b}~. \tag{A5}\label{vector prod}%
\end{equation}

In a matrix representation of the Pauli algebra $\mathcal{G}_{3}$ the basis
vectors $\vec{e}_{a}$ are represented by the Pauli matrices $\hat{\sigma}_{a}%
$, which we write as $\vec{e}_{a}\sim\hat{\sigma}_{a}$, and the geometric
product is represented by the matrix product. Then (\ref{Pauli a}) and
(\ref{pseudoscalar}) take the familiar forms
\begin{equation}
\hat{\sigma}_{a}\hat{\sigma}_{b}+\hat{\sigma}_{b}\hat{\sigma}_{a}=2\delta
_{ab}\hat{1}\quad\text{and}\quad\hat{\sigma}_{1}\hat{\sigma}_{2}\hat{\sigma
}_{3}=i\hat{1}~. \tag{A6}\label{Pauli b}%
\end{equation}

There are three involutions --- operations that when repeated yield the
identity. One is the operation of \emph{reversion}\ which consists of
reversing the order of all products of vectors. In the Pauli algebra
$\mathcal{G}_{3}$ this operation is denoted by a dagger $\dagger$ and is
defined by
\begin{equation}
(AB)^{\dagger}=B^{\dagger}A^{\dagger}\quad\text{and}\quad\vec{a}^{\dagger
}=\vec{a}~. \tag{A7}%
\end{equation}
In three dimensions we find that $i^{\dagger}=-i$ so that taking the reverse
of (\ref{multivector}), amounts to taking its Hermitian adjoint. The other
involution is the operation of \emph{spatial inversion} denoted by $\ast$ and
defined by
\begin{equation}
(AB)^{\ast}=A^{\ast}B^{\ast}\quad\text{and}\quad\vec{a}^{\ast}=-\vec{a}~.
\tag{A8}%
\end{equation}
In three dimensions this implies $i^{\ast}=-i$. The third involution is
obtained by taking both the reverse and the inverse, $A^{\dagger\ast}$.

We shall find it useful to take derivatives $\partial/\partial X=\partial_{X}$
with respect to a multivector $X$ \cite{Doran Lasenby 2003}\cite{Hestenes
Sobczyk 1984}. Just as the multivector $A$, eq.(\ref{multivector}), can be
expanded as
\begin{equation}
A=\langle A\rangle_{0}+\vec{e}_{a}\langle\vec{e}^{a\dagger}A\rangle_{0}%
+i\vec{e}_{a}\langle(i\vec{e}^{a})^{\dagger}A\rangle_{0}+i\langle i^{\dagger
}A\rangle_{0}~, \tag{A9}%
\end{equation}
where $\vec{e}^{a}=\delta^{ab}\vec{e}_{b}$ are the reciprocal basis vectors,
the multivector derivative $\partial_{X}$ with respect to
\begin{equation}
X=x_{0}+\vec{e}_{a}x^{a}+i\vec{e}_{a}y^{a}+iy_{0}~, \label{A10}%
\end{equation}
is given by its expansion in the reciprocal basis,
\begin{equation}
\partial_{X}=\frac{\partial}{\partial x_{0}}+\vec{e}^{a}\frac{\partial
}{\partial x^{a}}-i\vec{e}^{a}\frac{\partial}{\partial y^{a}}-i\frac{\partial
}{\partial y_{0}}~. \tag{A11}\label{multiv deriv b}%
\end{equation}
Most of the derivatives we shall need below can be computed using the Leibniz
rule for the derivative of a product and the formula
\begin{equation}
\partial_{X}\,\langle XA\rangle_{0}=\langle A\rangle_{X}~ \tag{A12}%
\label{multiv deriv c}%
\end{equation}
where $\langle A\rangle_{X}$ denotes the projection of $A$ onto those grades
contained in $X$.

\subsection*{A.2 Spinors}

Spinors in GA can be introduced as elements of a minimal left ideal
\cite{Riesz 1958} --- a subspace that remains invariant under multiplication
from the left and does not itself contain invariant subspaces; see
\cite{Hestenes 1966}. For example, a minimal left ideal of $\mathcal{G}_{3}$
is composed by multivectors $\Psi_{+}$ of the form
\begin{equation}
\Psi_{+}=\Psi\frac{1+\vec{e}_{3}}{2}~, \tag{A13}\label{spinor}%
\end{equation}
where $\Psi$ is an arbitrary multivector and $\left(  1+\vec{e}_{3}\right)
/2$ is a projector. Such elements will be called spinors. The set of elements
of the form $\Psi(1-\vec{e}_{3})/2$ constitute an independent left ideal.

A basis for the spinors $\Psi_{+}$ is given by
\begin{equation}
u_{+}=\frac{1}{\sqrt{2}}\left(  1+\vec{e}_{3}\right)  \quad\text{and}\quad
u_{-}=\vec{e}_{1}u_{+}=\frac{1}{\sqrt{2}}\vec{e}_{1}\left(  1+\vec{e}%
_{3}\right)  ~. \tag{A14}%
\end{equation}
The factors of $\sqrt{2}$ are chosen so that these basis spinors are
normalized according to
\begin{equation}
\langle u_{A}^{\dagger}u_{B}\rangle_{0}=\delta_{AB}~. \tag{A15}%
\label{spinor normalization}%
\end{equation}

In a matrix representation of $\mathcal{G}_{3}$, eq.(\ref{Pauli b}), the
matrices of a minimal left ideal are characterized by a single non-vanishing
column: using $\vec{e}_{k}\sim\hat{\sigma}_{k}$, we have%
\begin{equation}
u_{+}\sim%
\begin{bmatrix}
1 & 0\\
0 & 0
\end{bmatrix}
\quad\text{and}\quad u_{-}\sim%
\begin{bmatrix}
0 & 0\\
1 & 0
\end{bmatrix}
~. \tag{A16}%
\end{equation}
The action of $\vec{e}_{1}$, $\vec{e}_{2}$, and $\vec{e}_{3}$ on $u_{+}$ and
$u_{-}$ is precisely the action of the Pauli matrices $\hat{\sigma}_{1}$,
$\hat{\sigma}_{2}$, and $\hat{\sigma}_{3}$ on the eigenvectors of $\hat
{\sigma}_{3}$,
\begin{align}
\vec{e}_{1}u_{+}  &  =u_{-}\,;\quad\vec{e}_{2}u_{+}=iu_{-}\,;\quad\vec{e}%
_{3}u_{+}=u_{+}\,\nonumber\\
\vec{e}_{1}u_{-}  &  =u_{+}\,;\quad\vec{e}_{2}u_{-}=-iu_{+}\,;\quad\vec{e}%
_{3}u_{-}=-u_{-}~. \tag{A17}\label{spin a}%
\end{align}

\subsubsection*{A.2.1 Rotations and rotors}

Rotating a vector $\vec{r}$ by an angle $\theta$ about an axis described by
the unit vector $\vec{n}$ gives a new vector%
\begin{equation}
\vec{r}^{\prime}=U\vec{r}U^{\dag}~, \tag{A18}%
\end{equation}
where
\begin{equation}
U=\exp(-i\vec{n}\theta/2)=\cos\frac{\theta}{2}-i\vec{n}\sin\frac{\theta}{2}
\tag{A19}\label{rotor}%
\end{equation}
is a unitary multivector called a rotor. In the usual matrix representation,
$\vec{e}_{k}\sim\hat{\sigma}_{k}$, we have
\begin{equation}
U\sim\hat{U}=\exp(-in_{k}\hat{\sigma}_{k}\theta/2)~. \tag{A20}%
\end{equation}
Since $U$ and $-U$ represent the same rotation the\ rotors provide a
two-to-one representation of the rotation group, \emph{i.e.}, the rotors $U$
are elements of the corresponding \emph{spin group}, Spin(3). Furthermore,
since the grade of $i\vec{n}$ is even, the rotors $U$ belong to the even
subalgebra, $\mathcal{G}_{3}^{+}\overset{\text{def}}{=}\langle\mathcal{G}%
_{3}\rangle_{0+2}$, which is also known as the quaternion subalgebra. Thus,
there is a one-to-one correspondence between spinors in $\mathcal{G}_{3}$ and
quaternions. Indeed, using $\vec{e}_{3}u_{+}=u_{+}$ and (\ref{multivector}), a
generic spinor (\ref{spinor}) can be expressed as
\begin{equation}
\Psi=(\alpha+\vec{a}\,\vec{e}_{3}+i\vec{b}+i\vec{e}_{3}\beta)u_{+}~.
\tag{A21}\label{spinor a}%
\end{equation}
(From here on we shall drop the subscript $+$ and write $\Psi_{+}=\Psi$.) The
four terms in the parenthesis are of even grade and therefore, any spinor
$\Psi$ can be written as
\begin{equation}
\Psi=\Upsilon u_{+} \tag{A22}\label{spinor b}%
\end{equation}
where $\Upsilon$ is a quaternion. It will be useful to write the quaternion
$\Upsilon$ in \textquotedblleft polar\textquotedblright\ form, \
\begin{equation}
\Upsilon=\left\vert \Upsilon\right\vert \,U\quad\text{with}\quad UU^{\dag
}=U^{\dag}U=1~, \tag{A23}\label{spinor c}%
\end{equation}
where $\left\vert \Upsilon\right\vert =(\Upsilon^{\dag}\Upsilon)^{1/2}$ is the
magnitude and the unitary quaternion $U$ is a rotor. Since rotors $U$ act as
rotation operators, using eq.(\ref{spin a}) and (\ref{rotor}) we can see that
if the spinors $u_{+}$ and $u_{-}$ are eigenvectors of $\vec{e}_{3}$, then the
\textquotedblleft rotated\textquotedblright\ spinors $Uu_{+}$ and $Uu_{-}$ are
eigenvectors of the rotated vector $\vec{e}_{3}^{\,\prime}=U\vec{e}_{3}%
U^{\dag}$. Indeed,
\begin{align}
U\vec{e}_{3}u_{+}  &  =U\vec{e}_{3}U^{\dag}Uu_{+}=Uu_{+}\quad\text{or}%
\quad\vec{e}_{3}^{\,\prime}Uu_{+}=Uu_{+}~,\tag{A24}\\
\text{and\quad}U\vec{e}_{3}u_{-}  &  =U\vec{e}_{3}U^{\dag}Uu_{-}=-Uu_{-}%
\quad\text{or}\quad\vec{e}_{3}^{\,\prime}Uu_{-}=-Uu_{-}~. \tag{A25}%
\end{align}
Therefore a generic spinor such as $\Psi=\left\vert \Upsilon\right\vert
Uu_{+}$ is obtained by rotating $u_{+}$ by $U$ and dilating it by $\left\vert
\Upsilon\right\vert $.

\subsubsection*{A.2.2 Euler angle representation}

The orientation of an arbitrary frame $(x^{\prime},y^{\prime},z^{\prime})$
relative to a standard frame $(x,y,z)$ is specified by a rotor $U(\theta
,\phi,\chi)$ where $\{\theta,\phi,\chi\}$ are Euler angles. When expressed in
terms of rotations about the original standard frame axes the rotor
$U(\theta,\phi,\chi)$ is\footnote{The convention we adopt for the Euler angles
is common but by no means universal. Holland \cite{Holland 1993}, for example,
adopts a different convention.}
\begin{equation}
U(\theta,\phi,\chi)=U_{z}(\phi)U_{y}(\theta)U_{z}(\chi)=e^{-i\vec{e}_{3}%
\phi/2}e^{-i\vec{e}_{2}\theta/2}e^{-i\vec{e}_{3}\chi/2}~. \tag{A26}%
\label{U Euler a}%
\end{equation}
A useful expression for $U(\theta,\phi,\chi)\ $is obtained commuting the first
factor past the second to get
\begin{equation}
U(\theta,\phi,\chi)=e^{-i\vec{e}_{3}(\chi+\phi)/2}\cos\frac{\theta}{2}%
-i\vec{e}_{2}e^{-i\vec{e}_{3}(\chi-\phi)/2}\sin\frac{\theta}{2}~,
\tag{A27}\label{U Euler b}%
\end{equation}
which, using $-i\vec{e}_{2}u_{+}=u_{-}$, allows us to write a generic
normalized spinor $u$ as
\begin{equation}
u=U(\theta,\phi,\chi)u_{+}=u_{+}e^{-i(\chi+\phi)/2}\cos\frac{\theta}{2}%
+u_{-}e^{-i(\chi-\phi)/2}\sin\frac{\theta}{2}. \tag{A28}\label{spinor d}%
\end{equation}
The orthogonal spinor $u^{\prime}$ is
\begin{equation}
u^{\prime}=U(\theta,\phi,\chi)u_{-}=-u_{+}e^{i(\chi-\phi)/2}\sin\frac{\theta
}{2}+u_{-}e^{i(\chi+\phi)/2}\cos\frac{\theta}{2}. \tag{A29}\label{spinor e}%
\end{equation}

\subsection*{A.3 Rotors and the spin frame}

Let the three orthonormal vectors $\vec{e}_{1}$, $\vec{e}_{2}$, and $\vec
{e}_{3}$ be the standard basis for Euclidean space --- the laboratory basis.
Rotate the standard frame with different rotors $U(\vec{x})$ at different
points $\vec{x}$ in space. The result is a frame field $\{\vec{s}_{1},\vec
{s}_{2},\vec{s}_{3}\}$ with
\begin{equation}
\vec{s}_{k}(\vec{x})=U(\vec{x})\vec{e}_{k}U^{\dag}(\vec{x})~, \tag{A30}%
\end{equation}
which we will call the \emph{spin frame} or \emph{spin triad}. The orientation
of the spin frame relative to the standard frame $\{\vec{e}_{1},\vec{e}%
_{2},\vec{e}_{3}\}$ is specified through the local Euler angles $\{\theta
(\vec{x}),\phi(\vec{x}),\chi(\vec{x})\}$. From (\ref{U Euler a}) it is clear
that the angles $\theta$ and $\phi$ are the polar angles of $\vec{s}_{3}$
while the angle $\chi$ describes the rotation of $\vec{s}_{1}$ and $\vec
{s}_{2}$ about the $\vec{s}_{3}$ axis. From now on we drop the label $3$ and
the vector $\vec{s}_{3}=\vec{s}$ will be called the spin vector,
\begin{equation}
\vec{s}=U\vec{e}_{3}U^{\dag}=\sin\theta\cos\phi\,\vec{e}_{1}+\sin\theta
\sin\phi\,\vec{e}_{2}+\cos\theta\,\vec{e}_{3}~. \tag{A31}\label{spin vector}%
\end{equation}

\paragraph*{Infinitesimal rotations ---}

The effect of a change $\delta U$ in the rotor $U$ is to induce an additional
infinitesimal rotation by angles $\{\delta\theta,\delta\phi,\delta\chi\}$. To
see this start from $UU^{\dag}=1$ so that%
\begin{equation}
(\delta U)U^{\dag}+U\delta U^{\dag}=0\quad\text{or}\quad\delta U=-(U\delta
U^{\dag})U~. \tag{A32}%
\end{equation}
Next we note that $U\delta U^{\dag}$ is a bivector --- its grade is even ($U$
is even) and its scalar part $\langle U\delta U^{\dag}\rangle_{0}$ vanishes
because $(U\delta U^{\dag})^{\dagger}=-U\delta U^{\dag}$. Therefore we can
write $U\delta U^{\dag}$ as an imaginary vector,
\begin{equation}
U\delta U^{\dag}=i\delta\vec{\zeta}\quad\text{so that}\quad\delta
U=-i\delta\vec{\zeta}\,U~. \tag{A33}\label{dU zeta}%
\end{equation}
To find the meaning of $\delta\vec{\zeta}$ consider the rotor
\begin{equation}
U+\delta U\,=\left(  1-\frac{i}{2}\delta\vec{\zeta}\right)  U=e^{-\frac{i}%
{2}\delta\vec{\zeta}}U~. \tag{A34}%
\end{equation}
Thus, $\delta\vec{\zeta}$ represents an infinitesimal rotation by the angle
$|\delta\vec{\zeta}|$ about the axis $\vec{n}_{t}=\delta\vec{\zeta}%
/|\delta\vec{\zeta}|$. Using eq.(\ref{U Euler a}) to evaluate $(\delta
U)U^{\dagger}$ gives
\begin{equation}
\delta\vec{\zeta}=\vec{e}_{3}\delta\phi+\vec{e}_{\phi}\delta\theta+\vec
{s}\,\delta\chi~, \tag{A35}\label{dzeta a}%
\end{equation}
where
\begin{equation}
\vec{e}_{\phi}=-\sin\phi\,\vec{e}_{1}+\cos\phi\,\vec{e}_{2}~ \tag{A36}%
\end{equation}
and $\vec{s}$ is given in (\ref{spin vector}). Next we look at two special
cases of $\delta\vec{\zeta}$.

\paragraph*{Angular velocity ---}

The rotor and the spin frame can evolve over time and this change is
conveniently expressed in terms of an angular velocity vector $\vec{\omega
}_{t}$. The infinitesimal rotation described by
\begin{equation}
\delta\vec{\zeta}=\vec{\omega}_{t}\delta t\quad\text{where}\quad\vec{\omega
}_{t}=\vec{e}_{3}\partial_{t}\phi+\vec{e}_{\phi}\partial_{t}\theta+\vec
{s}\,\partial_{t}\chi~ \tag{A37}\label{omega t a}%
\end{equation}
leads to
\begin{equation}
\partial_{t}U=-\frac{i}{2}\vec{\omega}_{t}U~, \tag{A38}\label{omega t b}%
\end{equation}
and
\begin{equation}
\partial_{t}\vec{s}_{k}=\partial_{t}(U\,\vec{e}_{k}U^{\dag})=i\vec{s}%
_{k}\wedge\vec{\omega}_{t}=\vec{\omega}_{t}\times\vec{s}_{k}~. \tag{A39}%
\label{ds/dt}%
\end{equation}

\paragraph*{Spatial derivatives ---}

The spatial derivatives of the rotor $U$ and of the spin frame are
conveniently expressed by introducing three bivectors $\Omega_{a}$ and their
duals, the three vectors $\vec{\omega}_{a}$,%

\begin{equation}
\Omega_{a}=2U\partial_{a}U^{\dag}=-2(\partial_{a}U)U^{\dag}=i\vec{\omega}%
_{a}~, \tag{A40}\label{omega a}%
\end{equation}
where $\partial_{a}=\partial/\partial x^{a}$. The rotor at $\vec{x}+\delta
\vec{x}$ is
\begin{equation}
U(\vec{x}+\delta\vec{x})=U(\vec{x})+\delta x^{a}\partial_{a}U=\left(
1-\frac{i}{2}\vec{\omega}_{a}\delta x^{a}\right)  U(\vec{x})=e^{-\frac{i}%
{2}\vec{\omega}_{a}\delta x^{a}}U(\vec{x})~. \tag{A41}\label{dU omega t}%
\end{equation}
Thus, as one moves from $\vec{x}$ to $\vec{x}+\delta\vec{x}$ the spin frame is
rotated by the infinitesimal angle $|\vec{\omega}_{a}\delta x^{a}|$.
Therefore,
\begin{equation}
\delta\vec{\zeta}=\vec{\omega}_{a}\delta x^{a}\quad\text{where}\quad
\vec{\omega}_{a}=\vec{e}_{3}\partial_{a}\phi+\vec{e}_{\phi}\partial_{a}%
\theta+\vec{s}\,\partial_{a}\chi~,~ \tag{A42}\label{omega b}%
\end{equation}
which leads to
\begin{equation}
\partial_{a}U=-\frac{1}{2}\Omega_{a}U=-\frac{1}{2}i\vec{\omega}_{a}U~,
\tag{A43}\label{omega c}%
\end{equation}
and to the analogue of (\ref{ds/dt}),%

\begin{equation}
\partial_{a}\vec{s}_{k}=\vec{\omega}_{a}\times\vec{s}_{k}~.~ \tag{A44}%
\label{dsk}%
\end{equation}
The corresponding angles $\delta\zeta^{k}$ of rotation about the local axes
$\vec{s}_{k}$ are
\begin{equation}
\delta\zeta^{k}=\delta\vec{\zeta}\cdot\vec{s}_{k}~. \tag{A45}%
\end{equation}
Therefore,
\begin{equation}
\delta\vec{\zeta}=\delta\zeta^{k}\,\vec{s}_{k}\quad\text{and}\quad\vec{\omega
}_{a}=(\partial_{a}\zeta^{k})\vec{s}_{k}~. \tag{A46}\label{omega d}%
\end{equation}

The components of $\vec{\omega}_{t}$ and $\vec{\omega}_{a}$ along the spin
vector $\vec{s}$,
\begin{align}
\vec{\omega}_{t}\cdot\vec{s}  &  =\partial_{t}\zeta^{3}=\partial_{t}\chi
+\cos\theta\,\partial_{t}\phi~,\tag{A47}\label{w dot s a}\\
\vec{\omega}_{a}\cdot\vec{s}  &  =\partial_{a}\zeta^{3}=\partial_{a}\chi
+\cos\theta\,\partial_{a}\phi~, \tag{A48}\label{w dot s b}%
\end{align}
will play an important role; they will capture information about energy and
momentum. It is useful to write $\vec{\omega}_{t}\cdot\vec{s}$ and
$\vec{\omega}_{a}\cdot\vec{s}$ in terms of the spinor $u=Uu_{+}$,
\begin{equation}
\frac{1}{2}\vec{\omega}_{t}\cdot\vec{s}=\langle u^{\dagger}i\partial
_{t}u\rangle_{0}\quad\text{and}\quad\frac{1}{2}\vec{\omega}_{a}\cdot\vec
{s}=\langle u^{\dagger}i\partial_{a}u\rangle_{0}~. \tag{A49}\label{w dot s c}%
\end{equation}
The proof is straightforward. For example, using (\ref{spinor d}),
(\ref{omega b}), and the cyclic identity $\langle ABC\rangle_{0}=\langle
BCA\rangle_{0}$, we have
\begin{align}
\langle u^{\dagger}i\partial_{a}u\rangle_{0}  &  =\langle u_{+}U^{\dagger
}i\partial_{a}Uu_{+}\rangle_{0}=\langle(i\partial_{a}U)(1+\vec{e}%
_{3})U^{\dagger}\rangle_{0}\nonumber\\
&  =\frac{1}{2}\langle\vec{\omega}_{a}U(1+\vec{e}_{3})U^{\dagger}\rangle
_{0}=\frac{1}{2}\langle\vec{\omega}_{a}+\vec{\omega}_{a}\vec{s}\rangle
_{0}=\frac{1}{2}\vec{\omega}_{a}\cdot\vec{s}~. \tag{A50}\label{w dot s d}%
\end{align}

Beyond establishing the notation and a few useful results, the purpose of this
section has been to emphasize that the peculiar rotational properties of
spinors are of purely geometrical origin --- a far cry from any
\textquotedblleft peculiar, not classically describable
two-valuedness.\textquotedblright

\section*{Appendix B: HK flows imply linearity}

Equation (\ref{HK flow}) implies that $\tilde{Q}$ can be expressed as the sum
of scalar (grade zero) functionals
\begin{equation}
\tilde{Q}[\psi_{+x},\psi_{+x}^{\ast},\psi_{-x},\psi_{-x}^{\ast}]=\tilde{Q}%
_{0}+\tilde{Q}_{1}+\tilde{Q}_{2}+\tilde{Q}_{3}+\tilde{Q}_{4} \tag{B1}\label{Q}%
\end{equation}
where $\tilde{Q}_{0}$ is a constant, $\tilde{Q}_{1}$ is first order in $\psi
$s, $\tilde{Q}_{2}$ is quadratic in the $\psi$s, and so on. The functional
form of $\tilde{Q}$ can further be restricted as follows. First, since the
constant $\tilde{Q}_{0}$ does not contribute to the flow (\ref{H flow a}) it
can be set equal to zero, $\tilde{Q}_{0}=0$.

\paragraph*{The linear and cubic terms:}

The functionals $\tilde{Q}_{1}$ and $\tilde{Q}_{3}$ in (\ref{Q}) are ruled out
because they violate the global gauge symmetry (\ref{GGT}). This is
immediately clear once one writes
\begin{equation}
\tilde{Q}_{1}[\Psi,\Psi^{\dagger}]=\int dx\,\langle\Psi_{x}A_{x}^{\dagger
}+A_{x}\Psi_{x}^{\dagger}\rangle_{0}~ \tag{B2}%
\end{equation}
where $A$ is some generic multivector kernel and, similarly, for $\tilde
{Q}_{3}$ which involves integrands of the form%

\begin{equation}
\left\langle A\Psi B\Psi^{\dagger}C\Psi D\right\rangle _{0}\quad\text{or}%
\quad~\left\langle A\Psi^{\dagger}B\Psi C\Psi^{\dagger}D\right\rangle _{0}~.
\tag{B3}%
\end{equation}

\paragraph*{The quadratic term:}

The term $\tilde{Q}_{2}$ in (\ref{Q}) is bilinear in $\Psi$ and $\Psi
^{\dagger}$. Gauge invariance restricts us to functionals of the form
\begin{equation}
\tilde{Q}_{2}[\Psi,\Psi^{\dagger}]=\int dx_{1}dx_{2}\,\langle A_{x_{1}x_{2}%
}\Psi_{x_{2}}B_{x_{2}x_{1}}\Psi_{x_{1}}^{\dagger}\rangle_{0}~\, \tag{B4}%
\end{equation}
where $A$ and $B$ are multivector kernels. $\tilde{Q}_{2}$ can be simplified
as follows. Introducing a matrix representation,
\begin{align}
\Psi_{x_{1}}B_{x_{1}x_{2}}\Psi_{x_{2}}^{\dagger}  &  =%
\begin{bmatrix}
\psi_{+x_{1}} & 0\\
\psi_{-x_{1}} & 0
\end{bmatrix}%
\begin{bmatrix}
B_{11,x_{1}x_{2}} & B_{12,x_{1}x_{2}}\\
B_{21,x_{1}x_{2}} & B_{22,x_{1}x_{2}}%
\end{bmatrix}%
\begin{bmatrix}
\psi_{+x_{2}}^{\ast} & \psi_{-x_{2}}^{\ast}\\
0 & 0
\end{bmatrix}
\nonumber\\
&  =B_{11,x_{1}x_{2}}\Psi_{x_{1}}\Psi_{x_{2}}^{\dagger}~, \tag{B5}\label{B11}%
\end{align}
shows that the matrix elements $B_{12}$, $B_{21}$, and $B_{22}$ do not
contribute and can therefore be set to zero. The remaining matrix element
$B_{11,x_{1}x_{2}}$, which is just a complex number, can be absorbed into the
kernel $A_{x_{1}x_{2}}$. Let $B_{11}A=\hat{Q}$, then
\begin{equation}
\tilde{Q}_{2}[\Psi,\Psi^{\dagger}]=\int dx_{1}dx_{2}\langle\Psi_{x_{1}%
}^{\dagger}\hat{Q}_{x_{1}x_{2}}\Psi_{x_{2}}\rangle_{0}~.\, \tag{B6}%
\label{Q funct a}%
\end{equation}
Furthermore, since $\langle M^{\dagger}\rangle_{0}=\langle M\rangle_{0}$ for
any multivector $M$,\ we have
\begin{equation}
\tilde{Q}_{2}[\Psi,\Psi^{\dagger}]=\int dx_{1}dx_{2}\langle\Psi_{x_{2}%
}^{\dagger}Q_{x_{1}x_{2}}^{\dagger}\Psi_{x_{1}}\rangle_{0}=\int dx_{1}%
dx_{2}\langle\Psi_{x_{1}}^{\dagger}Q_{x_{2}x_{1}}^{\dagger}\Psi_{x_{2}}%
\rangle_{0} \tag{B7}%
\end{equation}
which implies that the kernel $A$ can be taken to be Hermitian,
\begin{equation}
\hat{Q}_{x_{2}x_{1}}^{\dagger}=\hat{Q}_{x_{1}x_{2}}~. \tag{B8}%
\label{Q hermitian}%
\end{equation}
More explicitly, we can introduce a matrix representation
\begin{equation}
\Psi_{x_{1}}^{\dagger}\hat{Q}_{x_{1}x_{2}}\Psi_{x_{2}}=%
\begin{bmatrix}
\psi_{+x_{1}}^{\ast} & \psi_{-x_{1}}^{\ast}\\
0 & 0
\end{bmatrix}%
\begin{bmatrix}
\hat{Q}_{11,x_{1}x_{2}} & \hat{Q}_{12,x_{1}x_{2}}\\
\hat{Q}_{21,x_{1}x_{2}} & \hat{Q}_{22,x_{1}x_{2}}%
\end{bmatrix}%
\begin{bmatrix}
\psi_{+x_{2}} & 0\\
\psi_{-x_{2}} & 0
\end{bmatrix}
~, \tag{B9}%
\end{equation}
and write the Hamiltonian functional as
\begin{align}
\tilde{Q}_{2}[\Psi,\Psi^{\dagger}]  &  =\int dx_{1}dx_{2}\,\langle\psi
_{+x_{1}}^{\ast}\hat{Q}_{11,x_{1}x_{2}}\psi_{+x_{2}}+\psi_{+x_{1}}^{\ast}%
\hat{Q}_{12,x_{1}x_{2}}\psi_{-x_{2}}\nonumber\\
&  +\psi_{-x_{1}}^{\ast}\hat{Q}_{21,x_{1}x_{2}}\psi_{+x_{2}}+\psi_{-x_{1}%
}^{\ast}\hat{Q}_{22,x_{1}x_{2}}\psi_{-x_{2}}\rangle_{0}~ \tag{B10}%
\label{Q funct b}%
\end{align}
which shows the various possible couplings between the spin up and down amplitudes.

\paragraph*{The quartic terms:}

The $\tilde{Q}_{4}$ functional takes the generic form
\begin{equation}
\tilde{Q}_{4}=\int dx_{1}dx_{2}dx_{3}dx_{4}\left\langle K_{x_{1}x_{2}%
x_{3}x_{4}}\psi_{+x_{1}}\psi_{+x_{2}}^{\ast}\psi_{-x_{3}}\psi_{-x_{4}}^{\ast
}\right\rangle _{0}~,~ \tag{B11}\label{Q4 a}%
\end{equation}
where the kernel $K$ is a complex number. Its action is to generate non-linear
HK flows that extend beyond the reach of the familiar unitary operators that
are provided within the linear framework of quantum mechanics. These quartic
functionals are ruled out because they are in conflict with rotational
invariance. Indeed, from (\ref{spinor d}) and (\ref{spinor e}) we see that
rotating the wave function $\Psi$, eq.(\ref{spinor psi d}), by Euler angles
$\{\theta^{\prime},\phi^{\prime},\chi^{\prime}\}$,
\begin{equation}
\Psi^{\prime}(x)=\psi_{+}^{\prime}(x)u_{+}+\psi_{-}^{\prime}(x)u_{-},
\tag{B12}%
\end{equation}
has the effect of mixing the amplitudes $\psi_{+}$ and $\psi_{-}$,
\begin{align}
\psi_{+}^{\prime}  &  =\psi_{+}e^{-i(\chi^{\prime}+\phi^{\prime})/2}\cos
\frac{\theta^{\prime}}{2}-\psi_{-}e^{i(\chi^{\prime}-\phi^{\prime})/2}%
\sin\frac{\theta^{\prime}}{2}~,\tag{B13}\\
\psi_{-}^{\prime}  &  =\psi_{+}e^{-i(\chi^{\prime}-\phi^{\prime})/2}\sin
\frac{\theta^{\prime}}{2}+\psi_{-}e^{i(\chi^{\prime}+\phi^{\prime})/2}%
\cos\frac{\theta^{\prime}}{2}~. \tag{B14}%
\end{align}
This means that a rotated $\tilde{Q}_{4}$ will include terms that violate the
HK condition (\ref{HK flow}). Note also that a rotation mixes $\psi_{+}$ and
$\psi_{-}$ but does not mix $\psi$ and $\psi^{\ast}$. Therefore a rotated
$\tilde{Q}_{2}$, eq.(\ref{Q funct b}), does not violate (\ref{HK flow}) while
a rotated $\tilde{Q}_{4}$ does. Thi concludes the derivation of
eq.(\ref{Q funct}).

\section*{Appendix C}

\label{App A} Here we derive the identity (\ref{identity}). Start from
eq.(\ref{drift c}) with $\gamma=1/2$ and use (\ref{w dot s a}) to write
\[
mv_{a}=-\frac{\hbar}{2}\vec{\omega}_{a}\cdot\vec{s}-\frac{q}{c}A_{a}~,
\]
so that
\begin{equation}
\frac{1}{2}mv^{2}=\frac{1}{2m}\left(  \frac{\hbar}{2}\vec{\omega}_{a}\cdot
\vec{s}+\frac{q}{c}A_{a}\right)  ^{2}~. \tag{C1}\label{A KE}%
\end{equation}
On the other hand, substituting $\Psi=\Upsilon u_{+}$\thinspace\ with
$\Upsilon=\rho^{1/2}U$ into (\ref{H0 a}) gives
\begin{equation}
(\hat{H}_{0}\Psi)\Psi^{\dag}=(\hat{H}_{0}\Upsilon)\Upsilon^{\dag}(1+\vec{s})~.
\tag{C2}\label{A Hpsipsi}%
\end{equation}
Next expand $\hat{H}_{0}\Upsilon$ to write
\begin{align}
\hat{H}_{0}\Upsilon &  =\left[  -\frac{\hbar^{2}}{2m}\partial^{2}%
\Upsilon+\frac{q^{2}}{2mc^{2}}A^{2}\Upsilon\right] \nonumber\\
&  +\left\{  i\frac{\hbar q}{2mc}(\partial_{a}A_{a})\Upsilon+i\frac{\hbar
q}{mc}(A_{a}\partial_{a})\Upsilon-\frac{\hbar q}{2mc}\vec{B}\Upsilon\right\}
~. \tag{C3}%
\end{align}
The terms in brackets are even; those in braces are odd. Therefore,
substituting back into (\ref{A Hpsipsi}) gives
\begin{align}
\langle\Psi^{\dag}\hat{H}_{0}\Psi\rangle_{0}  &  =\left\langle -\frac
{\hbar^{2}}{2m}(\partial^{2}\Upsilon)\Upsilon^{\dag}\right\rangle _{0}%
+\frac{q^{2}}{2mc^{2}}A^{2}\rho\nonumber\\
&  +\left\langle i\frac{\hbar q}{mc}A_{a}(\partial_{a}\Upsilon)\Upsilon^{\dag
}\vec{s}\right\rangle _{0}-\frac{\hbar q}{2mc}\vec{B}\cdot\vec{s}\rho~.
\tag{C4}\label{H0 b}%
\end{align}
Next use (\ref{omega c}) to write
\[
\partial_{a}\Upsilon=\frac{1}{2}(\frac{\partial_{a}\rho}{\rho}-i\vec{\omega
}_{a})\Upsilon~,
\]%
\[
\partial^{2}\Upsilon=[\frac{\partial^{2}\rho^{1/2}}{\rho^{1/2}}-\frac{1}%
{4}\vec{\omega}_{a}^{2}-\frac{i}{2}\partial_{a}\vec{\omega}_{a}-i\vec{\omega
}_{a}\frac{\partial_{a}\rho}{2\rho}]\Upsilon~.
\]
and substitute back into (\ref{H0 b}) to get
\begin{equation}
\langle\Psi^{\dag}\hat{H}_{0}\Psi\rangle_{0}=-\frac{\hbar^{2}}{2m}%
\frac{\partial^{2}\rho^{1/2}}{\rho^{1/2}}\rho+\frac{\hbar^{2}}{8m}\vec{\omega
}_{a}^{2}\rho+\frac{q^{2}}{2mc^{2}}A^{2}\rho+\frac{\hbar q}{2mc}A_{a}%
\vec{\omega}_{a}\cdot\vec{s}\rho-\frac{\hbar q}{2mc}\vec{B}\cdot\vec{s}\rho~.
\tag{C5}%
\end{equation}
Finally, use
\begin{equation}
\vec{\omega}_{a}^{2}=(\vec{\omega}_{a}\cdot\vec{s})^{2}+(\vec{\omega}%
_{a}\times\,\vec{s})^{2}=(\vec{\omega}_{a}\cdot\vec{s})^{2}+(\partial
_{a}\,\vec{s})^{2}~, \tag{C6}%
\end{equation}
and eq.(\ref{A KE}), to get
\begin{equation}
\langle\Psi^{\dag}\hat{H}_{0}\Psi\rangle_{0}=\rho\left(  \frac{1}{2}%
mv^{2}-\frac{\hbar^{2}}{2m}\frac{\partial^{2}\rho^{1/2}}{\rho^{1/2}}%
+\frac{\hbar^{2}}{8m}(\partial_{a}\,\vec{s})^{2}-\frac{\hbar q}{2mc}\vec
{B}\cdot\vec{s}\right)  ~, \tag{C7}%
\end{equation}
which concludes the derivation of (\ref{identity}).

\end{document}